\documentclass[pre,aps,floatfix,onecolumn,preprintnumbers,superscriptaddress,notitlepage,longbibliography]{revtex4-1}
\usepackage[utf8]{inputenc}
\usepackage[T1]{fontenc}

\usepackage{graphicx}
\usepackage{dcolumn}
\usepackage{bm}
\usepackage{amssymb}
\usepackage{pifont}
\usepackage{amsmath}
\usepackage{times}

\usepackage{epstopdf}
\usepackage{latexsym}
\usepackage{keyval}
\usepackage{ifthen}
\usepackage{moreverb}
\usepackage{color}

\usepackage{lipsum}

\newcommand{\av}[1]{\left\langle {#1} \right\rangle}
\newcommand{\by}{\overline{y}}

\DeclareMathOperator{\Real}{Re}

\graphicspath{{../},{SM/}}

\begin{document}
\title{Amplitude death and restoration in networks of oscillators with
random-walk diffusion}
\author{Pau Clusella}
\affiliation{Departament de F\'{\i}sica, Universitat Polit\`ecnica de
Catalunya, Campus Nord B4, 08034 Barcelona, Spain}
\author{M. Carmen Miguel}
\affiliation{Departament de F\'{\i}sica de la Mat\`{e}ria Condensada,
Universitat de Barcelona, Mart\'{\i} i Franqu\`es 1, 08028 Barcelona, Spain}
\affiliation{Universitat de Barcelona Institute of Complex Systems (UBICS),
Universitat de Barcelona, Barcelona, Spain}
\author{Romualdo Pastor-Satorras}
\affiliation{Departament de F\'{\i}sica, Universitat Polit\`ecnica de
Catalunya, Campus Nord B4, 08034 Barcelona, Spain}

\begin{abstract}
  We study the death and restoration of collective oscillations in networks
  of oscillators coupled through random-walk diffusion. Differently than the
  usual diffusion coupling used to model chemical reactions, here the
  equilibria of the uncoupled unit is not a solution of the coupled
  ensemble.  Instead, the connectivity modifies both, the original unstable
  fixed point and the stable limit-cycle, making them node-dependent.  Using
  numerical simulations in random networks we show that, in some cases, this
  diffusion induced heterogeneity stabilizes the initially unstable fixed
  point via a Hopf bifurcation. Further increasing the coupling strength the
  oscillations can be restored as well. Upon numerical analysis of the
  stability properties we conclude that this is a novel case of amplitude
  death. Finally we use a heterogeneous mean-field reduction of the system
  in order to proof the robustness of this phenomena upon increasing the
  system size.
\end{abstract}

\maketitle

\section{Introduction}

Complex systems composed by many units interrelated through a heterogeneous
topological pattern of interactions form a wide class of natural and man
made systems that can be fruitfully represented in terms of complex
networks~\cite{Newman10}.  Under this representation, in which nodes
stand for units and edges for pairwise unit interactions, a natural
framework arises that is capable to unify the functional and structural
properties of a wide variety of different systems.  Particularly important
at this respect are those systems that are the substrate for some dynamic of
transport process, whose properties can be strongly impacted by the topology
of interaction~\cite{BarratBarthelemyVespignani08,Dorogovtsev2008}.

A versatile formalism to describe dynamical processes on networks is given
by the theory of reaction-diffusion processes, described in terms of
different kinds of particles or species that diffuse along the edges of the
network and interact among them inside the nodes. Reaction-diffusion systems
find a natural representation in terms of sets of nonlinear differential
equations, representing in-node interactions, coupled by a diffusion term
representing transport between nodes. Successful applications of this
formalism can be found, for example, in the study of ecological species
dispersal~\cite{Hata:2014aa} or epidemic spreading~\cite{v.07:_react}. In
these studies, the network structure represents a
metapopulation~\cite{Hanski:2004}, defined as a group of populations or
physical patches, joined by migration or mobility paths.

In the context of reaction-diffusion systems on networks, it is
particularly noteworthy the work of Nakao and Mikhailov~\cite{Nakao2010},
where it was outlined the relation  between the usual understanding of
pattern formation in reaction-diffusion systems in lattices, established by
the seminal work of Turing~\cite{Turing52}, and its counterpart in irregular
topologies. In both cases, the stability of the homogeneous equilibrium,
preserved by the diffusion, can be established by means of the dispersion
relation, which, in the complex network case, relies on the diagonalization
of a Laplacian operator~\cite{Nakao2010, Asllani2014}. As a result, Turing
patterns are observed to arise in networked substrates, characterized by a
node-dependend pattern of species density~\cite{Nakao2010}. The observation of
Nakao and Mikhailov has spurred a flurry of activity in the field of
reaction-diffusion processes on networks, leading to the study of the
effects of directedness in the network edges~\cite{Asllani2014}, the
competition in predator-prey models~\cite{Fernandes2012},  effects on limit
cycles~\cite{Challenger2015}, collective synchronization~\cite{Cencetti2017}
or irregular Turing patterns~\cite{Cencetti2018}. 

Most previous works, however, are based on models where the coupling follows
Fick's diffusion law, used to model chemical reactions and heat transport.
In this case, the exchange rate of the physical quantities between two nodes
is proportional to their density difference.  A different choice for the
diffusive coupling that can be considered on complex networks corresponds to
random walk diffusion~\cite{v.07:_react,Baronchelli2008}. In this case the
exchange rate between two nodes is proportional to the inverse of their
degree, thus corresponding to particles diffusing by jumping between
randomly chosen nearest neighbor sites. This version of diffusion is
particularly relevant in the case of  ecology dynamics where each node
represents a metapopulation of a certain species in an ecosystem, which then
might randomly migrate to the surrounding
environments~\cite{turchin1998quantitative,okubo2013diffusion}.  Random
walks have been extensively studied in complex networks~\cite{Masuda2017},
but their application in the context of reaction-diffusion systems is rather
limited~\cite{v.07:_react,Baronchelli2008,CencettiLatora2018}. 

The differences between the two coupling schemes significantly affect the
nature of the system. As starting point, random-walk diffusion does not
accept, in general, a homogeneous equilibrium, hence the steady states are topology
dependent.  As direct implication of this situation, even if an equilibrium point
of the network is known, its stability cannot be attained by means of a
dispersion-relation. Overall, the emergent dynamical patterns that might
arise in reaction systems with random-walk diffusion have been hitherto
unexplored. In this paper we investigate one of such collective states which
is forbidden in reaction systems with Fick's diffusion: the quenching of the
oscillatory dynamics through the mechanism known as Amplitude Death (AD), as
opposed to the Oscillation Death (OD) mechanism that has been indeed studied
in systems with Fick's diffusion~\cite{Koseka2013}.

Oscillatory systems represent an important class for the choice of the
reaction terms.  Complex oscillatory systems are indeed used as a proxy to
study many relevant phenomena such as chemical
reactions~\cite{Kuramoto-84,Prigogine1968}, cardiac
cells~\cite{Guevara1981,Michaels1987} neural
dynamics~\cite{buzsaki2006rhythms,Izhikevich2008}, and ecological
fluctuations~\cite{Freund2006,Baurmann2007}.  Amplitude Death and
Oscillation Death are the two main routes through which the coupling among
the oscillatory units leads to a state of steadiness~\cite{Koseka2013}.
Although the two mechanisms have been prone to confusion, they correspond to
two different dynamical phenomena, with different implications on their
applicability.  The emergence of OD occurs when the coupling among units
induces the creation of new inhomogeneous stationary solutions. Upon
modifying the coupling strength, an originally oscillatory solution, such as
a limit-cycle, is destabilized and the system falls into the steady state
created by the coupling.  Nevertheless, the (unstable) oscillatory solution
and the inhomogeneous fixed point
coexist~\cite{Prigogine1968,Challenger2015}.  On the other hand, AD occurs
when different coupled oscillators pull each other out of the limit-cycle
upon increasing the interaction strength.  Thus, the amplitude of the
oscillations diminish until it completely vanishes and the oscillators fall
into the homogeneous fixed point of the system. Therefore, in this case, the
oscillatory solution collapses into the steady state and there is no
coexistence~\cite{Aronson1990,Mirollo1990,Reddy1998}.

In oscillatory systems coupled with Fick's diffusion it is possible to
compute the dispersion relation associated to the homogeneous time-varying
solution, whose instability leads to Turing patterns.  If these patterns are
steady in time, then we are before a case of OD~\cite{Challenger2015}.
Nevertheless, as we review in detail in this paper, the AD phenomena is
forbidden in networks with identical reaction terms and Fick's diffusion,
since the instability of the homogeneous fixed point is preserved by the
coupling~\cite{Aronson1990}.  Therefore one needs to invoke further
complexity on the description of the model, such as distributed
frequencies~\cite{Mirollo1990}, delayed interactions~\cite{Reddy1998} or
dynamic coupling~\cite{Konishi2003}.

Choosing random-walk diffusion strongly modifies this scenario.  Here we
show that an increase of the diffusion strength diminishes the amplitude of
the oscillations until they collapse into an inhomogeneous steady state.
This phenomena differs from OD in the sense that there is no coexistence
between the oscillatory solution and the fixed point.  We show that the
stationary solution corresponds to the uncoupled local equilibria of each
node that has been modified anisotropically by the coupling.  Therefore, the
cease of the oscillations corresponds to a novel case of AD where the
stabilized solution is inhomogeneous.  Here we extensively study this
transition towards AD and the later restoration of the oscillations, as well
as we show how suitable modifications of the network topology lead towards
the disappearance of AD.  We also perform heterogeneous mean-field analysis
in order to validate the generality of this phenomena for large systems.


\section{Gradient-driven diffusion}
\label{section1}

General two-species reaction-diffusion processes on a network can be
represented by the set of equations
\begin{equation*}
  \begin{cases}
	  \dot x_i=&f(x_i,y_i)+\displaystyle{D_x\sum_{j=1}^N \Delta_{ij}x_j}\\[10pt]
	  \dot y_i=&g(x_i,y_i)+\displaystyle{D_y\sum_{j=1}^N \Delta_{ij}y_j}\;
  \end{cases} ,
\end{equation*}
where  $D_x$ and $D_y$ are the coupling (diffusion) coefficients, $f$ and
$g$ are non-linear reactive terms, and the matrix $\Delta_{ij}$ is the
discrete Laplacian operator specifying the diffusive transport of species
between connected nodes. If diffusive transport is ruled by Fick's law, that
is, by the sum of fluxes of incoming species at each node, where the flux is
assumed to be proportional to the concentration~\cite{Samukhin2008}, the
discrete Laplacian can be written as $\Delta_{ij} = a_{ij} -
k_i\delta_{ij}$, where  $k_i$ is the degree of node $i$, $\delta_{ij}$ is
the Kronecker symbol and $a_{ij}$ is the network adjacency matrix, taking
value $1$ when nodes $i$ and $j$ are connected, and zero
otherwise~\cite{Newman10}.  In this case the equations of motion ruling the
dynamics of the $i$th node of the network read
\begin{equation}\label{eq1}
  \begin{cases}
  \dot x_i=&f(x_i,y_i)+\displaystyle{D_x\sum_{j=1}^N a_{ij}(x_j-x_i)}\\[10pt]
  \dot y_i=&g(x_i,y_i)+\displaystyle{D_y\sum_{j=1}^N a_{ij}(y_j-y_i)}
  \end{cases} .
\end{equation}

For this gradient-driven diffusion scheme, any solution of the uncoupled
system ($D_x=D_y=0$) corresponds to a solution of the coupled system.
Indeed, the diffusion term only depends on the density difference between
connected nodes, so if all nodes evolve with the exact same dynamics,
$(x_j(t),y_j(t))=(x(t),y(t))$ for all $j=1,\dots,N$, then the diffusive
coupling vanishes and the solution is preserved. Such solutions are
\emph{homogeneous}, meaning that the dynamic evolution of the system is the
same for all nodes independently of their topological properties. Since we
consider that the dynamics of each node has 2 dimensions,  only two types of
attractors are possible here: (i) fixed points of the original uncoupled
system, $(x^{(0)},y^{(0)})$, and (ii) periodic solutions, also referred to
as limit-cycles~\cite{strogatz2018nonlinear}.

\begin{figure*}[t]
  \centerline{\includegraphics[width=1\textwidth]{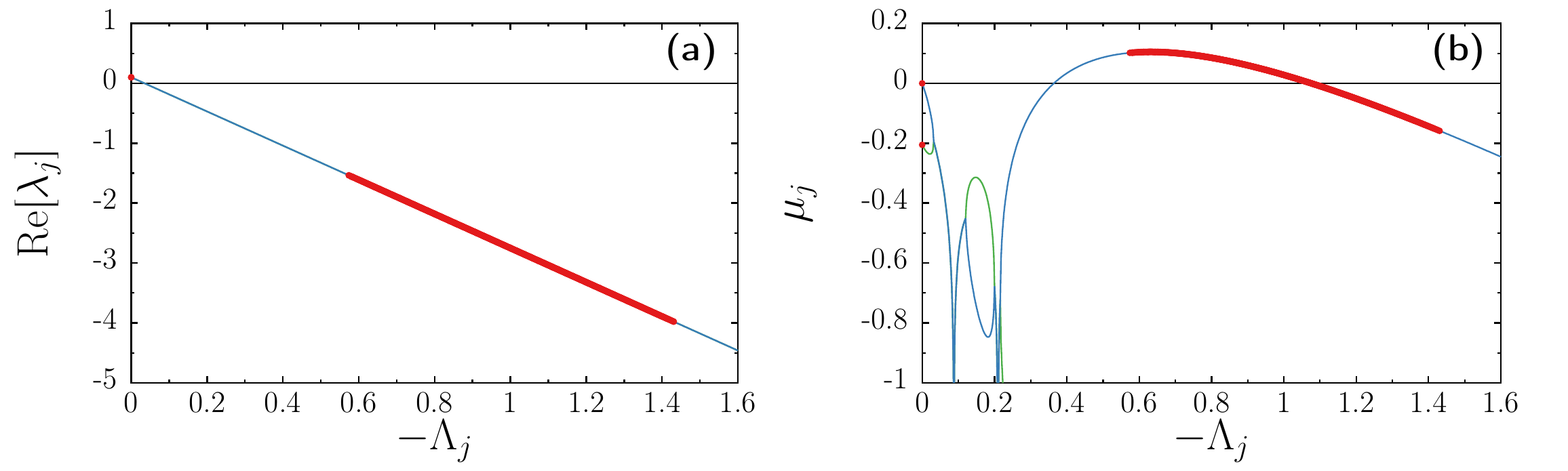}}
  \caption{Dispersion relation for the Brusselator system  coupled with
    Fick's diffusion on an Erd\"os–R\'enyi network with $N=1000$ and average
    degree $\langle k \rangle=20$. (a) Real part of the eigenvalues
    $\lambda_j$ controlling the stability of the homogeneous fixed point as
    a function of the Laplacian eigenvalues $\Lambda_j$. System parameters
    are $a=1.3$, $b=2.5$, $D_x=0.7$, and $D_y=5$ . (b) Real part of the
    Floquet exponents $\mu_j$ controlling the stability of the homogeneous
    limit-cycle solution as a function of the Laplacian eigenvalues
    $\Lambda_j$.  Continuous lines show the continuous dispersion relation,
	associated to a lattice at the thermodynamic limit (see Appendix \ref{app1}).
  Notice that with our definition of the Laplacian, the eigenvalues
$\Lambda_j$ are non-positive, thus we display $-\Lambda_j$ in the $x$-axis.
Other works might use a different definition.}
	\label{fig1}
\end{figure*}

The addition of the coupling diffusive terms can spontaneously modify the
stability of these homogeneous solutions. In principle, the stability of a
homogeneous fixed point of system (\ref{eq1}) boils down to study the
eigenvalues $\lambda_i$ of a $2N\times2N$ Jacobian matrix. Nevertheless it
is possible to relate the eigenvalues of such high-dimensional operator to
the eigenvalues of the Laplacian matrix of the system by means of a
dispersion relation
\begin{equation*}
  \lambda=F(\Lambda)
\end{equation*}
which maps all the Laplacian matrix eigenvalues $\Lambda_j$ to the
eigenvalues of the system Jacobian $\lambda_j$. This corresponds to the
extension of discrete dispersion-relations in lattices to the complex
network case \cite{Nakao2010,murray2006_vol2} (see Appendix \ref{app1} for a detailed
explanation).

In order to illustrate this situation, we consider as an example the
Brusselator model, whose dispersion relation can be worked out analytically
(see Appendix \ref{app2}). The resulting dispersion relation for specific system
parameters is depicted in Figure~\ref{fig1}(a), where we plot the real part
of the eigenvalues controlling the stability of $(x^{0},y^{(0)})$ as a
function of the eigenvalues of the Laplacian $\Lambda_j$ for the case of an
homogeneous  Erd\"os–R\'enyi network.  As it is well known, if
the network has a single connected component, the associated Laplacian
always contains a single zero eigenvalue corresponding to a uniform
eigenvector, and the rest are all negative.  In the dispersion relation,
such zero eigenvalue always returns the eigenvalues of the uncoupled
Jacobian (see Eq.~\eqref{eq:dispersion_general}). Thus, if the
fixed point of the original system is unstable, it will remain so despite
the coupling.  This is the case depicted in Figure~\ref{fig1}(a), where the
system eigenvalue corresponding to the uniform eigenvector has
$\Real[\lambda_0]=0.1$.  On the other hand, if the fixed point is originally
stable, the rest of the eigenvalues associated to the strictly negative
Laplacian eigenmodes might have a positive real part, thus destabilizing the
homogeneous solution and generating spatio-temporal patterns. This is the
well-known Turing instability, which triggers the so-called Turing
patterns~\cite{Turing52,murray2006_vol2,Nakao2010}.

This argument holds also for limit-cycles, where a numerical (and, in some
cases, analytical), dispersion relation can be obtained by means of Floquet
theory \cite{Challenger2015}.  Figure \ref{fig1}(b) depicts the relation
between the Floquet exponents $\mu_j$ of the limit-cycle for the Brusselator
system and the eigenvalues of the network Laplacian.  The Floquet exponents
corresponding to the uniform eigenvector are $\mu^{+}_0=0$ and
$\mu^{-}_0\simeq-0.216$, but there are many other Floquet exponents
associated to different network Laplacian modes that are positive. Thus, in
this case, the originally stable limit-cycle is being destabilized through a
Turing instability, so an arbitrary perturbation of the homogeneous solution
will develop heterogeneous patterns.  If these heterogeneous patterns are
stationary, then we would be before a case of OD \cite{Challenger2015}.
Regardless of whether this is the case or not, the AD phenomena will always
be forbidden here, since the instability of the original fixed point is
preserved when the coupling is added, and the also unstable oscillatory
regime still exists.

\section{Random walk diffusion}

Let us move now to a different coupling, the so called \emph{random-walk
diffusion}~\cite{Baronchelli2008,CencettiLatora2018}. Here variables $x_j$
and $y_j$ might represent the population density of animals that coexist in
an ecosystem divided in different discrete patches (a
metapopulation~\cite{Hanski:2004}). Animals interact inside patches
following some non-linear dynamics and can migrate between connected
patches. Representing the pattern of connections between patches by a
network, and assuming that animals move at random between patches, we have
that population of node $i$ diffuses uniformly through its $k_i$ neighbors,
each receiving a flux proportional to $1/k_i$.  The Laplacian matrix can
thus be written as $\tilde \Delta=(\tilde \Delta_{ij})$, with $\tilde
\Delta_{ij}=a_{ij}/k_j-\delta_{ij}$. The reaction-diffusion process then
reads 
\begin{equation}\label{eq:randomwalk1}
  \begin{cases}
    \dot x_i=&f(x_i,y_i)+D_x\displaystyle{\left( \sum_{j=1}^N a_{ij}
    \frac{x_j}{k_j}-x_i \right)} \\[10pt] \dot
    y_i=&g(x_i,y_i)+D_y\displaystyle{\left( \sum_{j=1}^N a_{ij}
    \frac{y_j}{k_j}-y_i \right)}\\
  \end{cases} .
\end{equation}
 
Here the solution of the uncoupled system ($D_x = D_y = 0$) is not a
solution of the coupled system unless we have a regular network, i.e. all
nodes have the same degree $k_i=k$ $\forall i$.  In this case, one can again
study the stability of the system by means of a dispersion relation and the
arguments exposed above still hold.  However, for generic non-regular
networks there is no theory that applies, and thus, the collective effects
induced by the coupling are unknown.

\begin{figure}[t]
  \centerline{\includegraphics[width=0.55\textwidth]{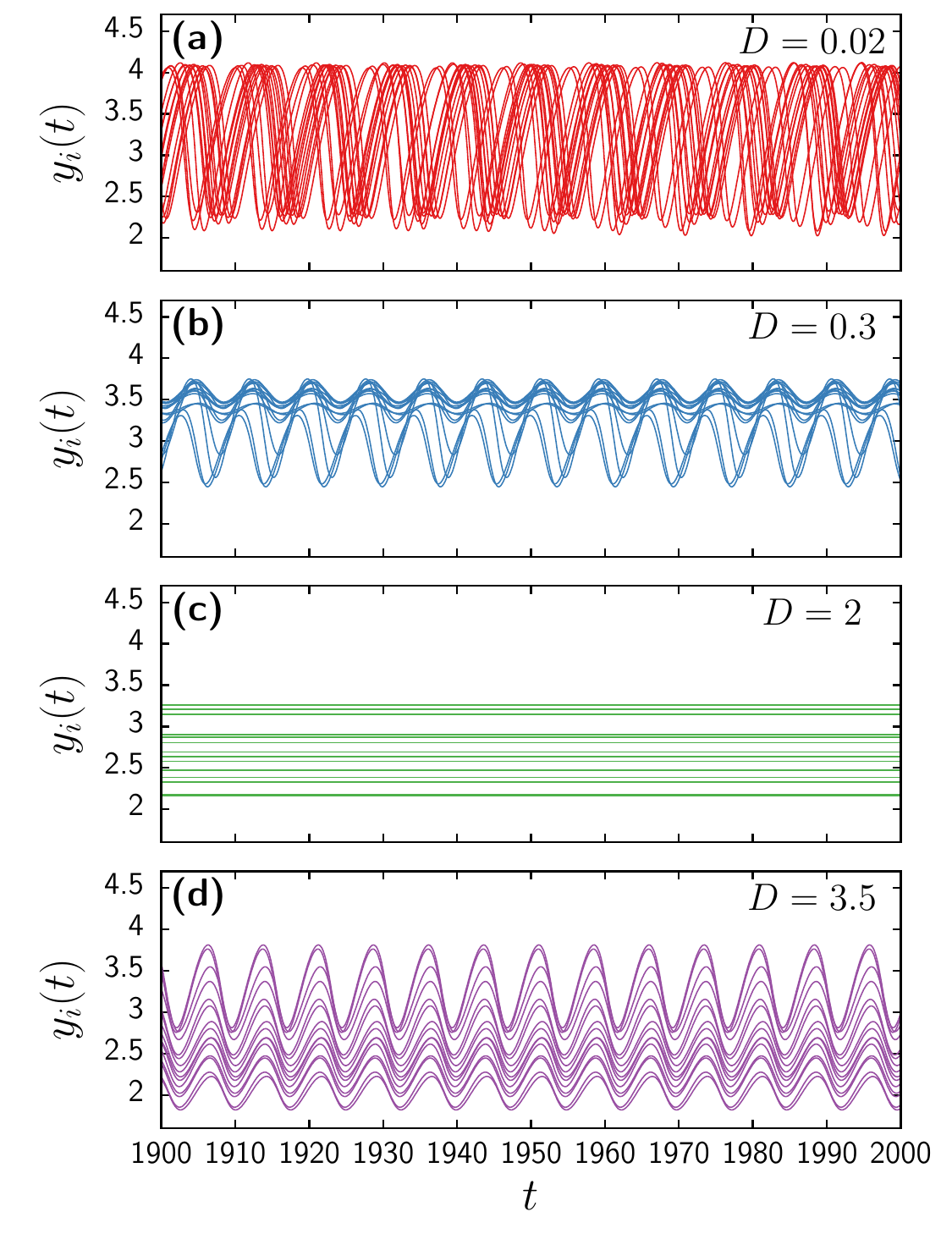}}
  \caption{Time series of the variable $y_i$ corresponding to 15 randomly
    chosen nodes (same nodes for each panel) obtained from the numerical
    integration of the Brusselator model Eq.~\eqref{eq:brusselator} with
    random walk diffusion Eq.~\eqref{eq:randomwalk1} on an Erd\"os–R\'enyi
    networks of size $N=1000$ and average degree $\langle k\rangle=20$.
    System parameters are $a=0.5$, $b=1.7$. The values of diffusion
  coefficient are (a) $D=0.02$, (b) $D=0.3$,(c) $D=2$, and (d) $D=3.5$.}
	\label{figTS}
\end{figure}

To study the effects of random walk diffusion, we consider the Brusselator
dynamics and integrate numerically the system defined by
Eq.~\eqref{eq:randomwalk1} in an Erd\"os–R\'enyi (ER) network with average
degree $\langle k\rangle=20$. For simplicity we set $D_x= D_y
= D$ throughout the rest of the paper. Unequal diffusion coefficients can
lead to much more complex dynamics; in fact, the generation of static Turing
patterns in models with Fick's diffusion requires $D_x\neq
D_y$~\cite{murray2006_vol2}.  In Figure~\ref{figTS} we show the evolution of
the variable $y_i$ resulting from numerical integration for different values
of the diffusion coefficient $D$. 

In Figure~\ref{figTS}(a) we can see that a very small diffusion $D=0.02$
does not strongly affect the behavior of the system. The limit cycle defined
by the variables in each node oscillates  with a very similar amplitude and
period, this last one close to the estimate for an isolated Brusselator, $T
\simeq 2 \pi/\sqrt{a}$. The different oscillators
fluctuate however out-of-phase.
Upon increasing the coupling,
(Figure~\ref{figTS}(b), corresponding to $D=0.3$), the oscillators evolve in
periodic phase-synchrony, but each having a different amplitude of the
limit-cycle, amplitude that is related to the degree of the corresponding
node in the network.  Increasing the coupling to  $D=2$,
(Figure~\ref{figTS}(c)), the system reaches a steady state, the heterogeneous
oscillations being substituted by an heterogeneous fixed point, whose value
is also related to the node's degree.  Further increasing of the coupling up
to $D=3.5$ (Figure~\ref{figTS}(d)) restores the in-phase synchrony, with
node dependent amplitude. These preliminary observations indicate the
existence of an oscillation quenching phenomenon, and a later oscillation
restoration. In the oscillatory regimes for $D>0$, the nodes appear to have
the same period (except in the vicinity of the first oscillation quenching
transition), period that depends on $D$, as a power spectrum analysis
reveals (see Appendix \ref{app7}).
Whether it corresponds to OD or AD is not yet clear.

In order to characterize the oscillation quenching behavior of the system we
consider the average of variable $y$ over the whole network,
\begin{equation*}
  \overline y(t) = \frac{1}{N}\sum_{j=1}^N y_j(t)\;,
\end{equation*}
and measure the temporal average $\langle \overline y \rangle$ and standard
deviation  $\sigma = \sigma(\overline y)$, defined as
\begin{equation}
  \langle \overline y \rangle = \lim_{T\to\infty} \int_0^T  \overline y(t')
  \; dt', \qquad \sigma(\by)^2 = \lim_{T\to\infty} \int_0^T  \overline y(t')^2
  \; dt' - \langle \overline y \rangle^2,
\end{equation}
that works as a measure the average amplitude of oscillations, with
$\sigma=0$ indicating a steady state.

In Figure.~\ref{fig:simulations}(a) we show in blue circles the value of
$\av{y}$ as a function of the diffusion coefficient $D$. The associated
error bars indicate the corresponding standard deviation $\sigma(\by)$. For
small values of $D$ the network displays oscillatory dynamics, as indicated
by the non-zero amplitude $\sigma(\by)$.  Upon increasing $D$, such
oscillations diminish until they vanish completely for diffusion larger than
$D_1\simeq 0.3775$. The system is frozen in a heterogeneous steady state
until the oscillations are restored again for a diffusion $D$ larger than
$D_2\simeq 3.095$.  We are therefore in front of  a case of oscillation
quenching occurring at $D_1$, followed by a subsequent oscillation
restoration when $D$ crosses $D_2$.  Whereas the inhomogeneity of the fixed
point would support the idea that the phenomenology corresponds to OD, the
two transitions between oscillatory and steady regimes strongly resemble the
AD phenomena, where the periodic solutions collapse into the fixed point,
which in this case happens to be heterogeneous as opposed to the common
cases of AD~\cite{Koseska2013,Koseka2013}.  In the folowing 
we study the two bifurcations through which the oscillations vanish and
restore in order to unveil the mechanism responsible for the cessation of
the oscillations.

\subsection{Heterogeneous fixed points}

\begin{figure*}[t]
  \centerline{\includegraphics[width=1\textwidth]{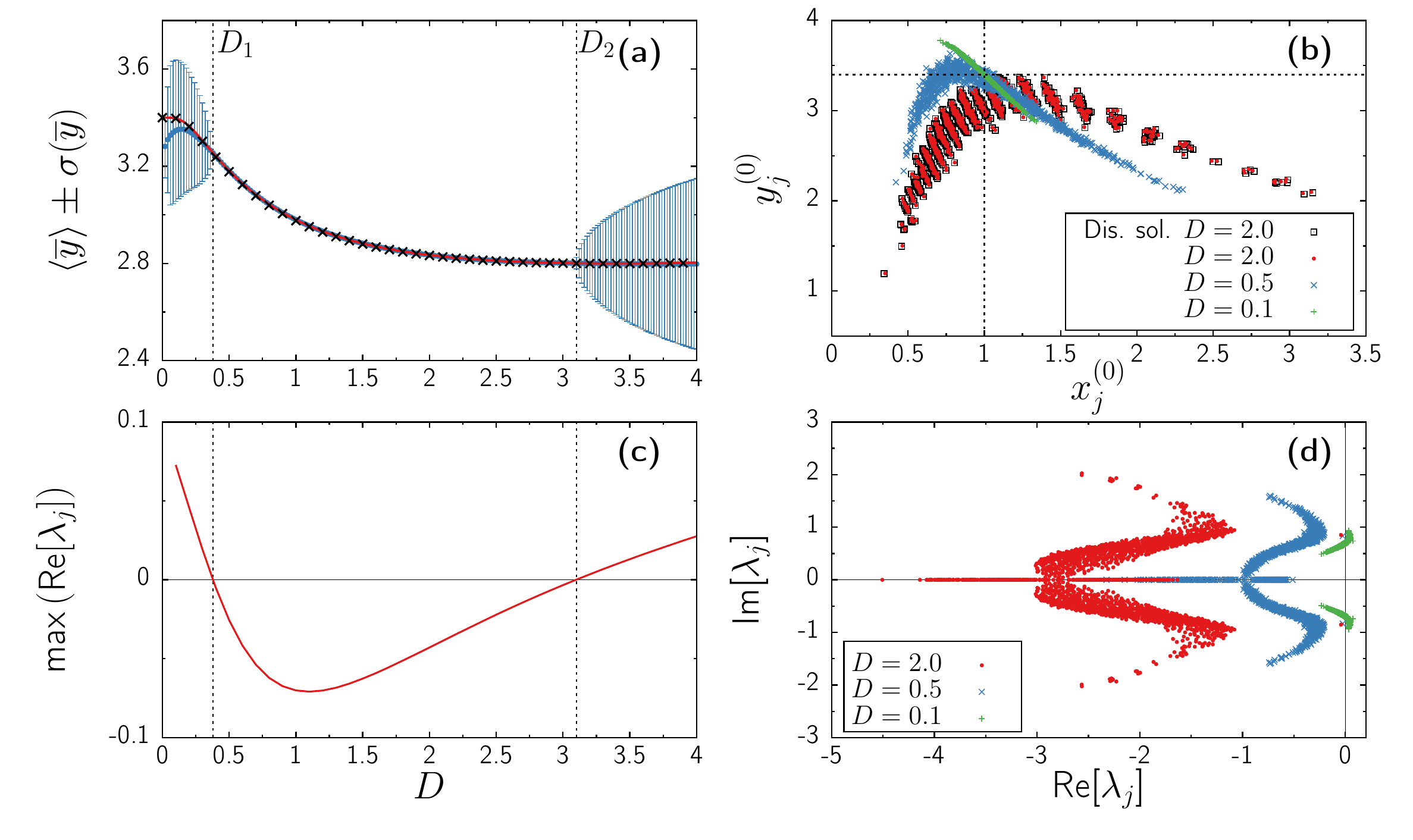}}
  \caption{Results of simulations and numerical analysis from the
    Brusselator model coupled through an ER network with $N=1000$ and
    average degree $\av{k}=20$. System parameters are $a=0.5$ and $b=1.7$.
    (a) Blue circles indicate the time-averaged variable $y$, $\langle
    \overline y \rangle$ obtained from numerical simulations for different
    values of diffusion $D$. Error bars indicate the standard deviation
    $\sigma(\overline y)$. Red continuous line shows the mean-field
    corresponding to the heterogeneous fixed point obtained solving
    numerically the system of equations (\ref{eq:fixedpoint}). (b)
    Heterogeneous fixed point. Green plusses, blue crosses, and red points
    correspond to numerical solutions of system (\ref{eq:fixedpoint}) for
    different values of $D$. Black squares correspond to the solution
    obtained by integrating Eq.(\ref{eq:dispersion}) up to $D=2$.The
    crossing of the two black dashed lines indicate the equilibria of the
  uncoupled system, $(x^{(0)},y^{(0)})=(1,b/a)$. (c) Largest eigenvalue's
real part for different values of $D$. (d) Eigenvalue spectra in the complex
plane for different values of $D$ (same symbols as in (b).}
	\label{fig:simulations}
\end{figure*}

In order to investigate the nature of these transitions, we start
considering the underlying fixed points, corresponding to the
solutions of the nonlinear system
\begin{equation}\label{eq:fixedpoint}
  \begin{cases}
    f(x_i,y_i)+\displaystyle{D\sum_{j=1}^N \tilde \Delta_{ij}x_j}=0,\\[10pt]
    g(x_i,y_i)+\displaystyle{D\sum_{j=1}^N \tilde \Delta_{ij}y_j}=0\;,
  \end{cases}
\end{equation}
that we solve numerically using a standard multidimensional root solver
based on the Newton-Raphson method. Figure~\ref{fig:simulations}(b) shows
the results obtained for different values of $D$ in the phase space.  For
any value of $D>0$, we obtain an inhomogeneous solution, i.e., depending on
the node $i$, that for small values of $D$ is close to the fixed point of
the uncoupled system $(x^{(0)}=1,y^{(0)}=b/a)$ (see green points). Upon
increasing $D$, the network equilibria spreads, covering a wider area of the
phase space.  For large diffusion (see for instance red dots, corresponding
to $D=2$)  the solution shows clusters of nodes with similar values which
correspond to nodes with the same degree.  It is worth noticing that we
find only one solution of Eq.~(\ref{eq:fixedpoint}) for each value of $D$.
Overall, it looks like the random-walk diffusion induces the hetoregeneity
of the equilibria, which becomes homogeneous only for $D=0$.

\subsection{Dispersion of the fixed point}

The previous numerical analysis points out that the heterogeneous
equilibrium of the network is linked  with the equilibrium of the uncoupled
system for small diffusion values, thus indicating that the fixed point of
the network corresponds to a coupling induced modification of the original
steady state.  We validate this assumption by exploring the effect that
small modifications of the  coupling strength $D$ have on the fixed point.
Let $(x_j^{(0)}(D),y_j^{(0)}(D))$ be the solution of Eqs.
(\ref{eq:fixedpoint}) for the diffusion value $D$. Assuming that the
dependence on $D$ is smooth, we consider a small increment of the diffusion,
$\epsilon>0$. Expanding up to first order terms in equation
(\ref{eq:fixedpoint}) one obtains (see Appendix \ref{app3} for a detailed derivation)
the system of equations
\begin{equation}\label{eq:dispersion}
  J\left(x_i^{(0)},y_i^{(0)}\right)
  \begin{pmatrix}
    \displaystyle{\frac{dx_i^{(0)}}{dD}}\\[10pt]
    \displaystyle{\frac{dy_i^{(0)}}{dD}}
  \end{pmatrix}
  +
  D\sum_{j=1}^N\tilde\Delta_{ij} 
  \begin{pmatrix}
    \displaystyle{\frac{dx_j^{(0)}}{dD}}\\[10pt]
    \displaystyle{\frac{dy_j^{(0)}}{dD}}
  \end{pmatrix}
  =-\sum_{j=1}^N\tilde\Delta_{ij}  \begin{pmatrix}
  x_j^{(0)}\\[10pt]
  y_j^{(0)}
  \end{pmatrix}
\end{equation}
for $i=1,\dots,N$, where $J(x,y)$ is the $2\times2$ Jacobian matrix of the
(uncoupled) reactive field, $(f(x,y) ,g(x,y))$.  Such equations form an
implicit linear non-autonomous system of differential equations with the
diffusion strength $D$ as independent variable.

An analytical solution of system in Eq.~(\ref{eq:dispersion}) is generally
unfeasible, but a simple numerical integration of such equation using
Euler's method, starting from the uncoupled system equilibrium, leads to the
average activity reported in black crosses in
Figure~\ref{fig:simulations}(a), whereas black squares in
Figure~\ref{fig:simulations}(b) correspond to the fixed point obtained with
this method for $D=2$. Both results match the solutions obtained by directly
solving Eqs. (\ref{eq:fixedpoint}), hence confirming that the heterogeneous
equilibria of the system coupled through random-walk diffusion corresponds
to a modification of the solution of the uncoupled system. In other words,
the steady state is not a completely new state induced by the coupling, but
a smooth transformation of the original equilibrium of the system, which
turns out to be node-dependent as soon as $D>0$.  Since the steady state
associated with OD corresponds to new states created by the coupling, the
observation that here the fixed point is not new is key to classify the
observed oscillation quenching mechanism as AD rather than OD.

\subsection{Stability analysis}

\begin{figure}[t]
  \centerline{\includegraphics[width=.5\textwidth]{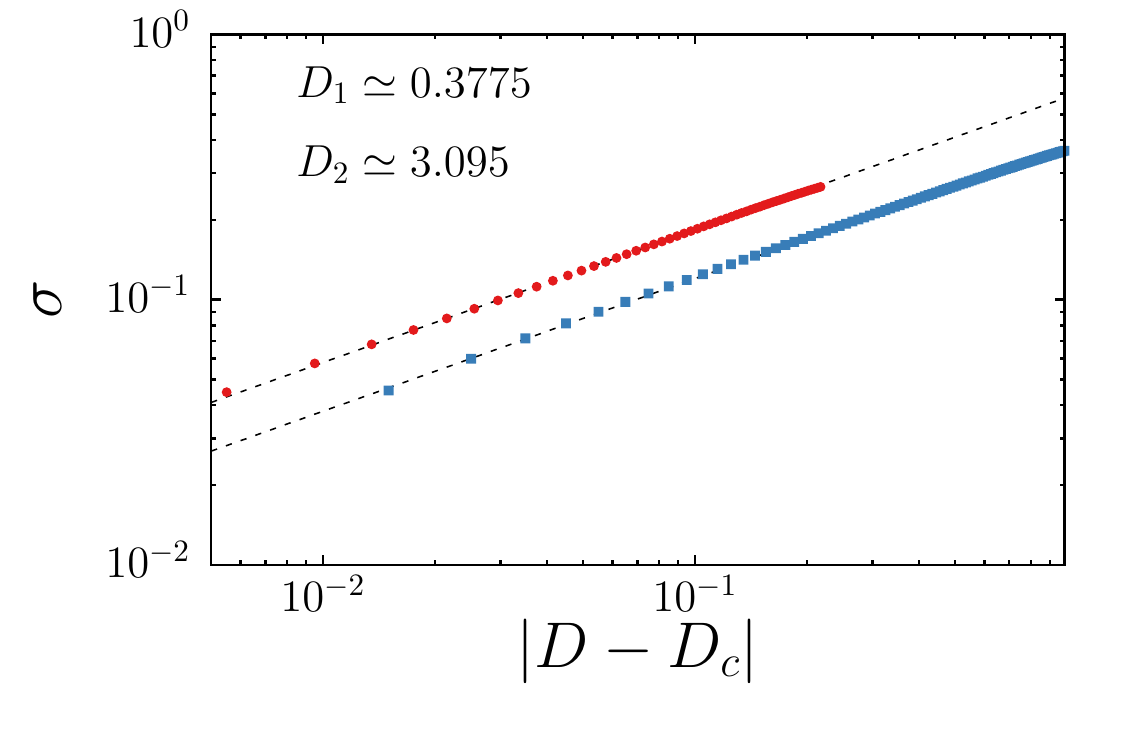}}
  \caption{Dependence of the oscillation amplitude $\sigma$ on $|D-D_1|$
  (red circles) and $|D-D_2|$ (blue squares). Black dashed lines correspond
to power laws with exponent $\beta=0.5$.}
	\label{fig:scaling}
\end{figure}

With the numerical solution of the system obtained by directly solving Eq.~
\eqref{eq:fixedpoint} one can study the stability of the system by
numerically computing the eigenvalues and eigenvectors of the full system
$2N\times2N$ Jacobian.  Figure~\ref{fig:simulations}(c) shows the dependence
of the real part of the maximum eigenvalue as the diffusion $D$ is tuned and
Figure~\ref{fig:simulations}(d) shows the full spectra for three different
values of the diffusion.  As it happens with the fixed point, for small $D$
all the eigenvalues are located close to the eigenvalues of the uncoupled
system $\lambda_\pm=0.1\pm0.7i$.  Thus, there is at least a pair of complex
conjugate eigenvalues with positive real part.  As $D$ increases, most of
the eigenvalues rapidly spread and are pushed to the left of the complex
plane. Nevertheless, a pair of complex conjugate eigenvalues remain isolated
from the rest and do not cross the imaginary axis until $D_1$ (see isolated
symbols close to the imaginary axis in figure~\ref{fig:simulations}(d)).
Upon further increasing the diffusion, the same isolated pair of eigenvalues
crosses again the imaginary axis at $D_2$, signaling the restoration of the
oscillations (see Fig.\ref{fig:simulations}(c)).  According to these results
it is clear that both transitions, $D_1$ and $D_2$, are Hopf supercritical
bifurcations.  Indeed, as shown in figure~\ref{fig:scaling}, the amplitude
of the oscillations vanishes as $\sigma\simeq\sqrt{|D-D_1|}$, as expected
for a supercritical Hopf bifurcation, and is restored then again at $D_2$
with the same exponent.  Also, the limit cycle centroid of each node
approaches the fixed point solution as $D$ approaches the bifurcation point
(see Appendix \ref{app7}).

A supercritical Hopf bifurcation induced by the coupling is one of the main
routes to Amplitude Death, the other being a saddle-node
bifurcation~\cite{Koseka2013}.  On the other hand, OD is associated to a
symmetry breaking pitchfork bifurcation, which allows for the coexistence of
the oscillatory solution and the steady state~\cite{Koseka2013}.  Here there
is no such symmetry breaking, the inhomogeneity of the fixed point being
induced by the heterogeneous coupling structure, that is not arbitrary.
Moreover, the limit-cycle perishes at the bifurcation points, and thus no
coexistence between the two solutions is possible.  This set of observations
altogether indicates that the oscillation quenching mechanism presented here
falls into the category of Amplitude Death, with the novelty that the
underlying fixed point is heterogeneous due to the combination of
random-walk diffusion and irregular network structure.  Although all the
results reported so far have been obtained using the Brusselator model, in
Appendix \ref{app6} we show the exact same phenomenology using the Holling-Tanner
predator-prey system~\cite{Tanner1975}.

\section{Sensitivity to topology and parameter modification}

\subsection{Network density}

In order to clarify the impact that irregular network topologies  have on
the stabilization of the fixed point and the later restoration of the
oscillations we have performed numerical simulations of the Brusselator
system Eq.~\eqref{eq:brusselator} with random-walk diffusion in different
classes of complex networks (in the Appendix \ref{app6} we report results corresponding to
different reaction terms).

Since an heterogeneous network topology is key for the emergence of the
amplitude death,  we first consider ER networks with different average
degree $\langle k\rangle$.  In order to avoid fluctuations due to the
independent generations of the topology, we construct  networks starting
form an initial ER configuration with given size and $\langle k\rangle=6$,
and progressively add at random new connections in order to generate denser
topologies with largest average degree.  Figures~\ref{fig:topoparam}(a,b)
show the amplitude of the mean-field oscillations for different $D$ using ER
networks with increasing average degree. In Figure~\ref{fig:topoparam}(a) we
can observe that AD takes place only in sparse networks, as increasing the
connectivity favors the oscillation amplitude.  In
Figure~\ref{fig:topoparam}(b) the black region corresponding to AD clearly
vanishes smoothly with increasing $\av{k}$.  In fact, for the smaller values
of the average degree, oscillations are not restored even for very large
values of $D$. As more connections are added
to the topology, the stable
steady state shrinks, until it completely vanishes for $\langle
k\rangle\simeq 45$.  The density of the connections thus is an important
factor to determine the existence of AD.

\subsection{Small-world network}

Apart from the overall connectivity density, the inner topological structure
of the network also plays a role in the AD and restoration.  Indeed, for
regular networks (where all nodes have the same degree), no quenching can
emerge, whereas irregularity seems to induce AD.  To unveil this situation,
we repeat the same analysis on small-world networks generated according to
the Watts-Strogatz (WS) algorithm~\cite{Watts1998} (see Methods). In this
case the network topology depends on a rewiring parameter $p\in[0,1]$. For
$p=0$ the generated architecture corresponds to a regular network and the
stability of the limit-cycle can be studied through the dispersion relation.
In this case, for $p=0$, no Turing-pattern is triggered and the fully
synchronized solution is the only stable attractor. Instead,  $p=1$
corresponds to a random network similar to the ER model, and the behavior of
the system does not differ much from the results reported in
figure~\ref{fig:simulations}(a).  Therefore, the behavior observed for
intermediate values of $p$ should reveal a topologically induced transition
from synchronization to amplitude death.

Figures~\ref{fig:topoparam}(c,d) show the outcome of such simulations in WS
networks for different values of $p$. To avoid fluctuations on the topology
generation algorithm, we construct the networks by progressively rewiring a
larger fraction of random edges in the same network. We use $1-p$ as
topological order parameter, measuring the network departure from randomness
, i.e., random for $1-p=0$ and regular for $1-p=1$.  As expected, for
$1-p=0$ the situation is similar to the ER topology with $\langle k
\rangle=20$: there is a first bifurcation towards steadiness for small $D$,
and a second bifurcation where oscillations are restored.  For larger
regularity, the two bifurcation points come closer together until they
collide and vanish for $p\simeq 0.36$.  Overall the scenario is analogous to
that seen in ER networks with increasing density, but now with the disorder
parameter $1-p$ playing the role of control parameter, instead of the
average degree.  Both scenarios are also reminiscent of the amplitude death
bifurcation observed in systems of oscillators with quenched
heterogeneity~\cite{Saxena2012,Mirollo1990}.  In our case, however, the
heterogeneity resides in the connectivity structure rather than in the
reactive terms, which remain homogeneous.

\subsection{System parameter}

\begin{figure*}[t]
  \includegraphics[width=1\textwidth]{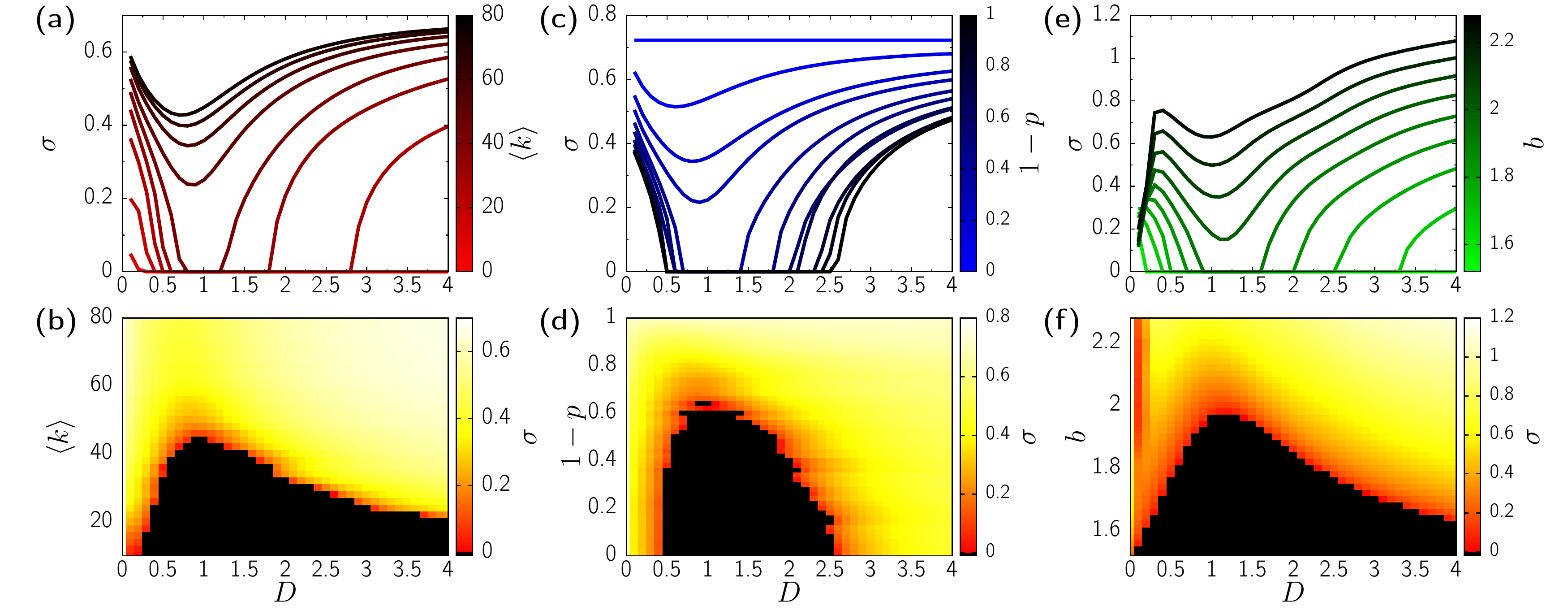} \caption{
    Amplitude death and restoration dependence on different topological and
    system parameters. Top panels show the dependence of $\sigma$ on $D$
    where each line corresponds to a specific topological or system
    parameter value. Bottom panels show heatmaps of the amplitude upon
    tuning the diffusion and the corresponding parameter. Results obtained
    with simulations of the Brusselator model in networks with $N=1000$ and
    system parameters $a=0.5$ and $b=1.7$ unless otherwise stated.  (a,b)
    Dependence of the oscillation amplitude $\sigma$ on the diffusion $D$ in
    a range of ER networks with different average degree $\overline k$.
    (c,d) Dependence of the oscillation amplitude $\sigma$ on the diffusion
    $D$ for a range of Watts-Strogatz networks with average degree $\av{k} =
    20$ and  rewiring probability $p\in[0,1]$.  (d).  (e,f) Dependence of
    the oscillation amplitude $\sigma$ on the diffusion $D$ for different
    values of the system parameter $b$ with fixed $0.5$ in a single ER
  networks with $\langle k\rangle=20$.}
	\label{fig:topoparam}
\end{figure*}

Finally, the choice of the dynamical parameters also plays a role on whether
amplitude death arises or not. Figure~\ref{fig:topoparam}(e,f) shows the
outcome of numerical simulations for the same ER network with $\av{k} = 20$,
keeping fixed the parameter $a=0.5$ and varying $b$. For $b < a+1 = 1.5$ the
reactive part does not display oscillations and the fixed point is stable.
For $b>a+1=1.5$ the uncoupled system undergoes a Hopf bifurcation and the
stable limit-cycle emerges.  Here we investigate the dynamics of the coupled
system for $b$ varying from $1.52$ to $2.1$. Not surprisingly, for values of
$b$ close to the bifurcation of the uncoupled system the amplitude death
regime is quickly attained at small values of $D$, and the restoration of
the oscillations is not triggered even for large coupling, i.e., the fixed
point stabilizes quickly with $D$.  As $b$ increases, the limit-cycle has a
larger amplitude and the diffusion-induced stabilization of the fixed point
requires larger diffusion values.  Indeed, the region of amplitude death
becomes smaller as they move away from  the single-node bifurcation point,
until it completely vanishes for $b \gtrsim 1.9$ in the same manner as it
does when the modifications are done in the topology of the network.

\section{Heterogeneous mean-field analysis}

In order to gain some insight on the behavior of amplitude death an
restoration induced by random walk diffusion, we apply the standard tool of
heterogeneous mean-field (HMF) theory~\cite{pv01a,Dorogovtsev2008},
specialized to reaction-diffusion processes~\cite{Baronchelli2008}.  The
basis of HMF consists in the annealed network
approximation~\cite{Dorogovtsev2008,Boguna09}, that replaces the static adjacency
matrix of a real network by an average over degree classes $\bar{a}_{ij}$
that, in the case of uncorrelated networks, takes the form
\begin{equation*}
  \bar{a}_{ij}=\frac{k_i k_j}{\av{k} N}\;.
\end{equation*}
Introducing this expression into Eq.~(\ref{eq:randomwalk1}), we obtain the
HMF dynamical equations
\begin{equation}\label{eq:meanfield}
\begin{cases}
\dot x_i=&f(x_i,y_i)+D\left( \tilde k_i \overline x-x_i \right) \\[10pt]
\dot y_i=&g(x_i,y_i)+D\left( \tilde k_i \overline y-y_i \right)\\
\end{cases}
\end{equation}
where  $\tilde k_i = k_i / \av{k}$, $\overline x=\frac{1}{N}\sum_{j=1}^N
x_j$ and $\overline y=\frac{1}{N}\sum_{j=1}^N y_j$ are the average
(mean-field) activities of $x$ and $y$ variables.  Within this framework,
the dynamics of each node depends only on its own degree,  the mean-field
values $\overline{x}$ and $\overline{y}$, and the average connectivity of the network.
In fact, one can assume then that all nodes with the same degree behave
identically, thus formally reducing the $2N$-dimensional system
(\ref{eq:meanfield}) to a $2n$-dimensional system where $n$ is the number of
different degrees in the network~\cite{Dorogovtsev2008}. An additional
interest of this approach lies in the fact that it allows to consider
network sizes much larger than those permitted in a direct numerical
integration, since in general $n \ll N$.

System (\ref{eq:meanfield}) can be solved semi-analytically assuming that
$\overline x$ and $\overline y$ are  parameters that are to be determined
self-consistently \emph{a posteriori} (see Appendix \ref{app5} for details).  Figure
\ref{fig:mf} shows the results of the heterogeneous mean-field
approximation.  In Figure~\ref{fig:mf}(a) we show the fixed point obtained
for the reduced system (black squares) compared to the actual fixed point
obtained solving Eq. (\ref{eq:fixedpoint}) (red circles) for $D=2$.  The
solution of the reduced system matches the overall spreading of the
solution, with the black squares fitting nicely the center of the bands
displayed by the actual fixed point, corresponding to nodes with the same
degree. In Figure~\ref{fig:mf}(b) the mean-field activity of $y$ determined
by the HFM equations (dashed black curve) matches with good accuracy that of
the actual system (red continuous curve). Moreover, the stability analysis
of the fixed point also coincides with that of the original system, with a
cloud of eigenvalues lying far on the stable part, and a pair of complex
conjugate eigenvalues crossing back and forth the imaginary axis upon
modifying $D$.  Thus, the mean-field reduction does not only provide a good
proxy to study the fixed point, but also allows to study the amplitude death
phenomena in terms of stability.

\subsection{Large network size limit}

\begin{figure*}[t]
	\includegraphics[width=1\textwidth]{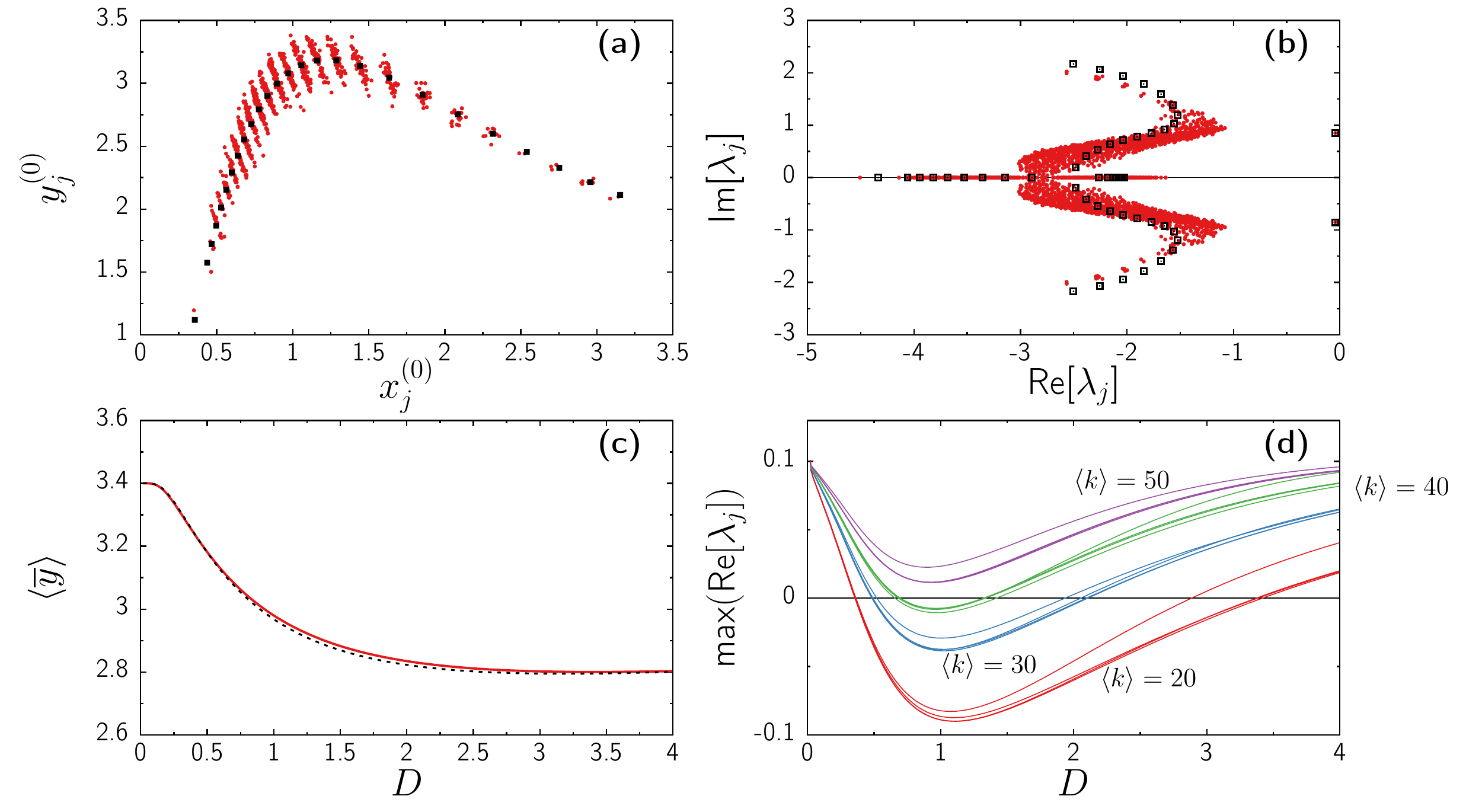} \caption{
		Mean-field analysis of the Brusselator model.  (a) Red circles
		correspond to the fixed point for $D=2$ as obtained from solving system
		(\ref{eq:fixedpoint}). Black squares correspond to the mean-field result
		for the same $D$.  (b) Red circles correspond to the eigenvalues of the
		original fixed point for $D=2$, whereas open black square indicate the spectra
		resulting from the mean-field reduction.  (c) Average activity of the
		fixed point obtained from directly solving system (\ref{eq:fixedpoint})
		(red continuous curve), and from the mean-field reduction (black dashed
		curve).  (d) Largest eigenvalue's real part for the mean-field solution
		with ER networks with average degree $\langle  k\rangle =30$ (red), 30 (blue),
		40 (green), and 50 (purple).  For each set of networks each line denotes
		a different network size. From top to bottom, $N=10^3$, $10^4$, $10^5$,
		and $10^6$.  }
	\label{fig:mf}
\end{figure*}

The good agreement of the mean-field theory with the numerical results of
the ER networks allows to extend our analysis to larger systems.  Figure
\ref{fig:mf}(d) shows the largest eigenvalue real part for different values
of $D$ as obtained from the heterogeneous mean-field analysis.  We analyzed
4 sets of ER networks of size  $N=10^3$, $10^4$, $10^5$, and $10^6$, with
different average degrees $\av{k} = 20$, $30$, $40$, and $50$. For a
specific value of $\langle k\rangle$, we observe that networks with
different sizes produce qualitatively similar results, converging to a well
defined limit for sufficiently large $N$. For instance, for $\av{k}=20$ the
upper red curve, corresponding to $N=10^3$, shows a smaller region of
amplitude death, whereas the red curves below, corresponding to larger
systems, converge nicely upon increasing $N$.  The same situation is
repeated for the other values of $\av{k}$, thus confirming that the effects
of the average degree on the amplitude death are not finite-size but rather
robust.  Indeed, the overall conclusion of this analysis is that the network
behavior depends strongly on the degree of each node, but is only mildly
dependent on system size. In the Appendix \ref{app7} we extend this analysis to the other
topological parameters displayed in Figure~\ref{fig:topoparam}, and in Appendix \ref{app6} to the
Holling-Tanner system.

\section{Discussion}

Reaction-diffusion processes are a powerful formalism to represent general
dynamical processes on networks, in which particles or species interact
inside nodes while moving diffusively between pairs of connected nodes. By
analogy with chemical reactions, a gradient-driven diffusion term, given by
Fick's law, is usually assumed. However, in certain circumstances, a random
walk diffusion term, in which particles jump at random along edges, might be
more realistic. Here we have shown how the nature of the diffusion term can
alter the behavior of the limit cycles in oscillatory reaction-diffusion
processes when driven by the diffusion term. Thus, while gradient-diffusion
can quench the oscillations by means of an
oscillation death mechanism, in which the stability of the original fixed
point is preserved by the diffusion operator, random walk diffusion
generates an amplitude death quenching characterized by a inhomogeneous
steady state induced by the structure of the diffusion operator. In this
case, a further increase in diffusion restores the oscillations, in the form of
a set of limit cycles with different amplitude for each node. The transitions
to amplitude quenching and restoration are observed to correspond to Hopf
supercritical bifurcations, in agreement with an amplitude death mechanism at
work. Our observations, backed up by direct numerical integration of the
Brusselator systems in random networks, are confirmed by a linear
stability analysis of the eigenvalues of the diffusion operator and by a
seminumerical heterogeneous mean-field approximation, which allows us to
check the phenomenology in very large networks. The role of random walk
diffusion in the amplitude quenching of oscillations, which is observed for
different classes of homogeneous networks, different sets of parameters and
even different reaction terms, is thus a robust phenomenology,
which could be relevant in processes such as epidemic spreading or
ecological dispersion. Future promising work along these lines would include
studying the effects of an heterogeneous, scale-free
topology~\cite{Barabasi:1999}, as observed in many natural networks.

\appendix
\label{section:methods}

\section{Dispersion relation in reaction-diffusion processes}
\label{app1}

For a detailed introduction on Turing patterns and how to compute
dispersion-relations on regular lattices the reader can check Murray's
book \cite{murray2006_vol2}. Here we summarize the results of Nakao
\textit{et al.} \cite{Nakao2010} on reaction-diffusion systems in complex
networks.  Let us consider a reaction-diffusion system in a complex network
with a gradient-driven diffusion term
\begin{equation*}
  \begin{cases}
  \dot x_i=&f(x_i,y_i)+\displaystyle{D_x\sum_{j=1}^N \Delta_{ij}x_j},\\[10pt]
  \dot y_i=&g(x_i,y_i)+\displaystyle{D_y\sum_{j=1}^N \Delta_{ij}y_j},
  \end{cases}
\end{equation*}
with a Laplacian matrix $\Delta_{ij} = a_{ij} - k_i\delta_{ij}$. This
Laplacian matrix is semi-definite negative, and its eigenvalues are all real
and non-positive.  Let $(x^{(0)},y^{(0)})$ be a fixed point of the uncoupled
system ($D_x = D_y = 0$), and hence the homogeneous solution of the coupled
system.  Inserting an arbitrary perturbation $\{(\delta x_i(t),\delta
y_i(t))\}_{i=1}^N$ of the homogeneous solution to the equations and
retaining up to first order terms one gets that the evolution of the
perturbation behaves as
\begin{equation}\label{eq:disprel}
  \begin{pmatrix}
    \displaystyle{\dot {\delta x_i}}\\[10pt]
    \displaystyle{\dot {\delta y_i}}
  \end{pmatrix}
    =J\left(x^{(0)}, y^{(0)}\right)
  \begin{pmatrix}
    \displaystyle{\delta x_i}\\[10pt]
    \displaystyle{\delta y_i}
  \end{pmatrix}
  +
  \sum_{j=1}^N \Delta_{ij}
  \begin{pmatrix}
    D_x\displaystyle{\delta x_j}\\[10pt]
    D_y\displaystyle{\delta y_j}
  \end{pmatrix}\;,
\end{equation}
where $J$ is the Jacobian of the homogeneous system, evaluated at
$(x^{(0)},y^{(0)})$.  Let
$\Phi^{(\alpha)}=(\phi_1^{(\alpha)},\dots,\phi_N^{(\alpha)})^T$ be the
Laplacian normalized eigenvector associated the eigenvalue
$\Lambda_\alpha$  for $\alpha=1,\dots,N$, such that
\begin{equation}\label{eq:diag}
\sum_m \Delta_{jm} \phi_m^{(\alpha)}=\Lambda_\alpha\phi_j^{(\alpha)}\;.
\end{equation}
We can express the perturbation of the homogeneous solution in terms of the
basis of the eigenvectors as
\begin{equation*}
  \begin{pmatrix}
    \displaystyle{\delta x_j}\\[10pt]
    \displaystyle{\delta y_j}
  \end{pmatrix}
=
\sum_{\alpha=1}^{N}  
  \begin{pmatrix}
    \displaystyle{ u^{(\alpha)}}\\[10pt]
    \displaystyle{ v^{(\alpha)}}
  \end{pmatrix}
  \phi_j^{(\alpha)}\;.
\end{equation*}
Applying this change of coordinates on Eq.~\eqref{eq:disprel} and making use
of the relation (\ref{eq:diag}) and of the linear independence of the
eigenvectors $\{\Phi^{(\alpha)}\}_{\alpha=1}^N$ one obtains that the
evolution of $(\dot{u}^{(\alpha)},\dot{v}^{(\alpha)})^T$ becomes independent
for each $\alpha=1,\dots,N$ through the relation
\begin{equation}
  \begin{pmatrix}
    \displaystyle{ \dot{u}^{(\alpha)}}\\[10pt]
    \displaystyle{ \dot{v}^{(\alpha)}}
\end{pmatrix}
=
J\left(x^{(0)}, y^{(0)}\right)
  \begin{pmatrix}
    \displaystyle{ u^{(\alpha)}}\\[10pt]
    \displaystyle{ v^{(\alpha)}} 
  \end{pmatrix}
+
\Lambda_\alpha
  \begin{pmatrix}
    \displaystyle{ D_x u^{(\alpha)}}\\[10pt]
    \displaystyle{ D_y v^{(\alpha)}}
  \end{pmatrix}
=
  \begin{pmatrix}
    \displaystyle{\frac{\partial f}{\partial x}\left(x^{(0)},y^{(0)}\right)}
    + D_x\Lambda_\alpha & \displaystyle{\frac{\partial g}{\partial
    x}\left(x^{(0)},y^{(0)}\right)} \\[10pt] 
    \displaystyle{\frac{\partial f}{\partial y}\left(x^{(0)},y^{(0)}\right)}
    & \displaystyle{\frac{\partial g}{\partial
    y}\left(x^{(0)},y^{(0)}\right)} + D_y\Lambda_\alpha
  \end{pmatrix} 
  \begin{pmatrix}
    \displaystyle{ u^{(\alpha)}}\\[10pt]
    \displaystyle{ v^{(\alpha)}}
  \end{pmatrix}\;.
  \label{eq:dispersion_general}
\end{equation}
Thus the stability of the homogeneous solution simplifies to the study of
the eigenvalues (and eigenvectors) of the previous $2\times 2$ matrix, which
are obtained as functions of the Laplacian eigenvalues $\Lambda_\alpha$. See
below for an explicit application to the Brusselator system.  The extension
of this analysis to limit-cycles makes use of Floquet theory,
see~\cite{Challenger2015}.

\section{The Brusselator}
\label{app2}

The Brusselator~\cite{Prigogine1968} is a prototypical model of
autocatalytic chemical reaction with two species showing oscillatory
behavior in the form of a limit cycle.  In its dimensionless form, in the
absence of diffusion, it is defined by the reaction terms
\begin{eqnarray}\label{eq:brusselator}
  f(x, y) &=& 1-x(b+1)+ax^2y,\\
  g(x, y) &=& bx-ax^2y, \nonumber
\end{eqnarray}
where $a,b>0$ are model  parameters.  The system has a single fixed point at
$(x^{(0)},y^{(0)})=(1,b/a)$, whose stability is ruled by the Jacobian
\begin{equation*}
  J\left(x^{(0)},y^{(0)}\right)=
  \begin{pmatrix}
    b-1  & a \\ -b & -a
  \end{pmatrix}\;.
\end{equation*}
The fixed point is therefore stable for  $b<a+1$.  Otherwise, the system
exhibits periodic behavior, with a period, for $b$ close to $1+a$,
approximately equal to $T = 2 \pi / \sqrt{a}$. The transition from the fixed
point to the oscillatory regime is through a supercritical Hopf bifurcation.
Unless otherwise stated, the system parameter values used in the paper are
$a=0.5$ and $b=1.7$.

When considering the Brusselator reaction terms (\ref{eq:brusselator}) with
Fick's diffusion (\ref{eq1}) one can obtain the dispersion relation of the
homogeneous fixed point.  Following Eq. (\ref{eq:dispersion_general}), one
only needs to to compute the eigenvalues of
\begin{equation*}
   \begin{pmatrix}
   b-1+D_x\Lambda_\alpha  & a \\[10pt] -b & -a+D_y\Lambda_\alpha
   \end{pmatrix}\;,
\end{equation*}
which yields
\begin{equation*}
\lambda_\pm(\Lambda_\alpha)=
\frac{1}{2}\left(b-a-1-\Lambda_\alpha(D_x+D_y)\pm\sqrt{(D_y-D_x)\Lambda_\alpha+b-1+a-4ab}\right).
\end{equation*}
This is the continuous curve displayed in Figure \ref{fig1}(a).

\section{Dispersion of the fixed point}
\label{app3}
Let $\left(x_i^{(0)}(D),y_i^{(0)}(D)\right)$ be the fixed point solution of 
the reaction-diffusion system (\ref{eq:randomwalk1}) with diffusion value $D$, so that
\begin{equation}\label{eq:methods1}
 \begin{cases}
 f\left(x_i^{(0)}(D),y_i^{(0)}(D)\right)+\displaystyle{D\sum_{j=1}^N \tilde \Delta_{ij}x_j^{(0)}(D)}=0\\[10pt]
 g\left(x_i^{(0)}(D),y_i^{(0)}(D)\right)+\displaystyle{D\sum_{j=1}^N \tilde \Delta_{ij}y_j^{(0)}(D)}=0\;.
 \end{cases}
\end{equation}
for $i=1,\dots,N$.  Assuming a smooth dependence of
$(x_j^{(0)}(D),y_j^{(0)}(D))$ on $D$ for $j=1,\dots,N$, we consider a small
increment of the diffusion $\epsilon>0$.  The aim is to solve then
\begin{align*}
  f\left(x_i^{(0)}(D+\epsilon),y_i^{(0)}(D+\epsilon)\right)&+\displaystyle{(D+\epsilon)\sum_{j=1}^N \tilde \Delta_{ij}x_j^{(0)}(D+\epsilon)}=0 \quad\text{and}\\
  g\left(x_i^{(0)}(D+\epsilon),y_i^{(0)}(D+\epsilon)\right)&+\displaystyle{(D+\epsilon)\sum_{j=1}^N \tilde \Delta_{ij}y_j^{(0)}(D+\epsilon)}=0\;.
\end{align*}
Expanding such equations around $(x_j^{(0)}(D),y_j^{(0)}(D))$ and retaining the first-order terms
one obtains, 
\begin{align*}
\epsilon \; f_x\left(x_i^{(0)}(D),y_i^{(0)}(D)\right)\frac{d x_i^{(0)}(D)}{d D}&
+\epsilon \; f_y\left(x_i^{(0)}(D),y_i^{(0)}(D)\right)\frac{d y_i^{(0)}(D)}{d D}
+\displaystyle{\epsilon\sum_{j=1}^N \tilde \Delta_{ij}\left(D\frac{d x_j^{(0)}(D)}{d D}+x_j^{(0)}(D)\right)}=0 \quad\text{and}\\
\epsilon \; g_x\left(x_i^{(0)}(D),y_i^{(0)}(D)\right)\frac{d x_i^{(0)}(D)}{d D}&
+\epsilon \; g_y\left(x_i^{(0)}(D),y_i^{(0)}(D)\right)\frac{d y_i^{(0)}(D)}{d D}
+\displaystyle{\epsilon\sum_{j=1}^N \tilde \Delta_{ij}\left(D\frac{d y_j^{(0)}(D)}{d D}+y_j^{(0)}(D)\right)}\;.
\end{align*}
Dividing then both sides of these equations by $\epsilon$
one finally can write the differential system that rules the dependence
of $(x_i^{(0)},y_i^{(0)})$ on $D$,
\begin{equation*}
J\left(x_i^{(0)},y_i^{(0)}\right)
\begin{pmatrix}
\displaystyle{\frac{dx_i^{(0)}}{dD}}\\[10pt]
\displaystyle{\frac{dy_i^{(0)}}{dD}}
\end{pmatrix}
+
D\sum_{j=1}^N\tilde\Delta_{ij} 
\begin{pmatrix}
\displaystyle{\frac{dx_j^{(0)}}{dD}}\\[10pt]
\displaystyle{\frac{dy_j^{(0)}}{dD}}
\end{pmatrix}
=-\sum_{j=1}^N\tilde\Delta_{ij}  \begin{pmatrix}
x_j^{(0)}\\[10pt]
y_j^{(0)}
\end{pmatrix}\qquad\text{for}\qquad i=1,\dots,N
\end{equation*}
where $J(x,y)$ is the $2\times2$ Jacobian matrix of the (uncoupled) reactive field, $(f(x,y) ,g(x,y))$.
The previous equation is thus an implicit linear non-autonomous system of differential equations
that can be solved by means of usual numerical integrators using as
initial condition for $D=0$ the solution of the uncoupled oscillators.

\section{Network models}
\label{app4}

\subsection{Erd\"os–R\'enyi networks}

The Erd\"os–R\'enyi (ER) model for network generation provides random
network topologies~\cite{Newman10}.  In particular we use the $G(N,p)$ model,
where a each pair of nodes is connected with probability $p\in[0,1]$. The
average degree of the network is then $pN$ and the degree distribution
follows a binomial distribution with parameters $N-1$ and $p$.

\subsection{Watts-Strogatz model}

The Watts-Strogatz (WS) model generates networks with small-world
connectivity, i.e., with small density and diameter, while still showing a
large degree of transitivity or clustering~\cite{Watts1998}. We use this
model as way to generate networks halfway between random and regular
topologies.  The model starts with $N$ nodes distributed on a ring, each
node being connected to its $\langle k\rangle$ nearest neighbors.  Then,
every edge of the network is rewired with probability $p\in[0,1]$.  The
small-world network class is represented for small values of the rewiring
probability, whereas for $p$ close to 1 the topology becomes closer to that
of a ER network.

\section{Heterogeneous mean-field analysis}
\label{app5}

We consider the HMF equations Eq.~(\ref{eq:meanfield}) specialized for the
Brusselator reaction terms. Solving for the fixed
point one obtains that (dropping the subindices for simplicity), we obtain
\begin{equation}\label{mfy}
y=\frac{1-x(b+1)+D(\tilde k \overline x -x)}{-ax^2}=\frac{bx+D\overline y}{ax^2+D}\;,
\end{equation}
from where one can obtain a cubic equation for the value of $x$:
\begin{equation*}
-a(D+1) x^3 + a(1+D(\overline x + \overline y))x^2-D(b+1+D)x+D(1+D\overline x)=0\;.
\end{equation*}
Thus 
\begin{equation}\label{mfx}
x=\frac{-1}{3a} \left( b+C + \frac{\Delta_0}{C} \right)
\end{equation}
where
\begin{align*}
C&=\left(\frac{\Delta_1\pm\sqrt{\Delta_1^2-4\Delta_0^3}}{2}\right)^{1/3}\\
\Delta_0&= a^2(1+D(\overline x+\overline y))^2-3aD(D+1)(b+1+D)\\
\Delta_1&=2a^2(1+D(\overline x + \overline y))^2 \\ &\qquad -9a^2D(D+1)(b+1+D)(1+D(\overline x+\overline y))\\ &\qquad+ 27 a^2(D+1)D(1+D\overline x)\;.
\end{align*}
The Brusselator system has the property that, in the fixed point and even
without using any kind of approximation, $\overline x=1$.  However, in order
to obtain $\overline y$ self-consistently one needs to rely on numerics.  In
practice, one starts with an educated guess for the mean-field $\overline y$
and determines then all $x_j$ and $y_j$ using equations (\ref{mfx}) and
(\ref{mfy}). Using a bisection method one can reduce the error between the
new estimated values $\overline x$ and $\overline y$ and the initial ones to
the desired accuracy.

\section{The Holling-Tanner predator-prey system}
\label{app6}

Here we revisit the results exposed on the main manuscript using a different
system for the single-node dynamics.  We use the Holling-Tanner
predator-prey system~\cite{Tanner1975}.  In this model $x\geq0$ and $y\geq0$
respectively represent the population fraction of a prey and predator
species in an ecosystem.  On its dimensionless form, the corresponding
reactive terms read
\begin{equation*}
\begin{cases}
f(x,y)&=x(1-x)-\displaystyle{\frac{axy}{x+d}}\\
g(x,y)&=yb \displaystyle{ \left( 1-\frac{y}{x} \right)}
\end{cases},
\end{equation*}
where $a,b$, and $d$ are system parameters~\cite{Murray2011}.

The system has a fixed point at
\begin{equation*}
x^{(0)}=\frac{1}{2}\left((1-a-d)+\sqrt{(1-a-d)^2+4d}\right),\qquad y^{(0)}=x^{(0)}
\end{equation*}
and the corresponding Jacobian is
\begin{equation*}
J\left(x^{(0)},y^{(0)}\right)=
\begin{pmatrix}
\displaystyle{x^{(0)}\left(\frac{ax^{(0)}}{(x^{(0)}+d)^2}-1\right)} & \displaystyle{\frac{-ax^{(0)}}{x^{(0)}+d}}\\[15pt]
b & -b
\end{pmatrix}\;.
\end{equation*}
The resulting eigenvalues have positive real part if and only if
\begin{equation*}
\displaystyle{x^{(0)}\left(\frac{ax^{(0)}}{(x^{(0)}+d)^2}-1\right)}<b\qquad\text{and}\qquad
1+\displaystyle{\frac{a}{u^{(0)}+d}}-\frac{ax^{(0)}}{(x^{(0)}+d)^2}>0\;.
\end{equation*}
Fixing $a=1$ and $d=0.1$, the fixed point becomes unstable through a
supercritical Hopf bifurcation at $b_c\simeq0.2625...$, so that for $b<b_c$
the single node dynamics exhibit periodic oscillations.  Unless otherwise
stated, we consider here  $b=0.25$.

\subsection{Amplitude death and restoration}

Now we consider a network of Holling-Tanner oscillators coupled with
random-walk diffusion.  Let $A=(a_{ij})$ be the adjacency matrix of the
(symmetric and fully connected) network, and let $k_j$ be the degree of node
$j$.  The random-walk Laplacian matrix is then defined as $\tilde
\Delta:=(\tilde \Delta_{ij})$  where $\tilde
\Delta_{ij}=a_{ij}/k_j-\delta_{ij}$.  With this notation, the dynamical rule
governing the evolution of the $i$th node of the system reads
 \begin{equation}\label{appeq1}
 \begin{cases}
 \dot x_i=&f(x_i,y_i)+\displaystyle{D\sum_{j=1}^N \tilde \Delta_{ij}x_j}\\[10pt]
 \dot y_i=&g(x_i,y_i)+\displaystyle{D\sum_{j=1}^N \tilde \Delta_{ij}y_j}
 \end{cases}
 \end{equation}
 where $D$ is the diffusion or coupling strength.\\

\begin{figure*}[t]
        \centerline{\includegraphics[width=1\textwidth]{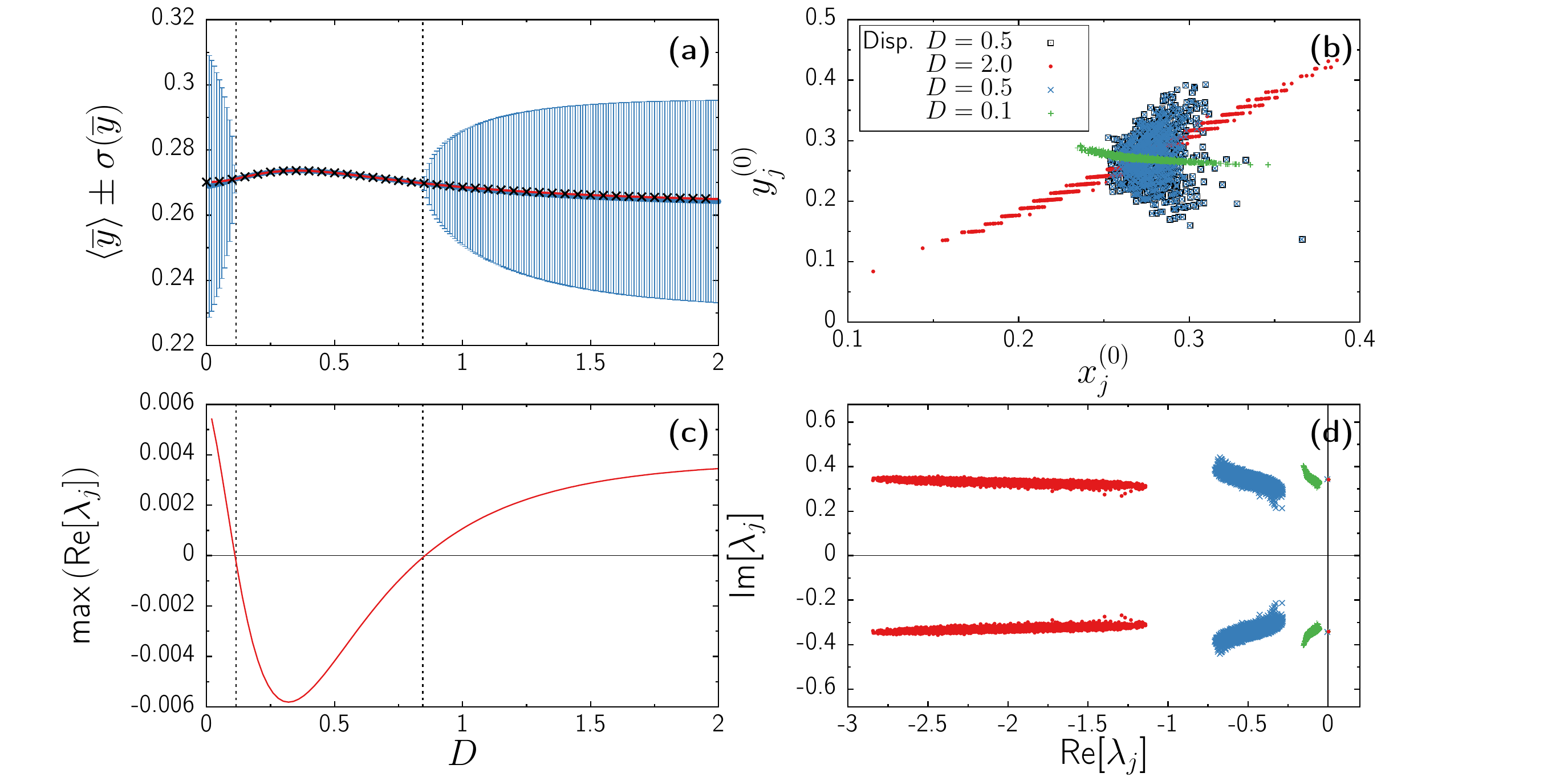}}
        \caption{ Results of simulations and numerical analysis from the
                Holling-Tanner model coupled through an Erdős–Rényi network with $N=1000$
                and average degree $\langle k\rangle=20$. System parameters are $a=1$, $b=0.25$, and
                $d=0.1$.  (a) Blue circles indicate the time-averaged mean-field
                activity of variable $y$, $\langle \overline y \rangle$ obtained from
                numerical simulations for different values of diffusion $D$. Error bars
                indicate the standard deviation $\sigma(\overline y)$. Red continuous
                line shows the mean-field corresponding to the heterogeneous fixed point
                obtained solving numerically the corresponding system of nonlinear equations. Black crosses indicate the same mean-field, but as obtained from the dispersion equation.
                 (b) Heterogeneous fixed point as obtained numerically solving the system of nonlinear equations for different values of $D$ (see legend). Black squares mark the result obtained by numerical integration of the dispersion equation. (c) Largest eigenvalue's
                real part for different values of $D$. (d) Eigenvalue spectra in the
                complex plane for different values of $D$ (symbols as in (b)).  }
        \label{appfig1}
\end{figure*}

Figure \ref{appfig1}(a) shows numerical results obtained using this system,
analogous to Figure~2 of the main manuscript. Overall, the scenario is akin
to the one exposed with the Brusselator model. The error bars of the blue
points in Figure~\ref{appfig1}(a) indicate the amplitude of the mean-field
oscillations, which decrease until become steady at $D_1\simeq 0.12$.
Further increasing the coupling, the oscillations restore around $D_2\simeq
0.86$.  Again, solving for the fixed point of the system with a
Newton-Raphson method one obtains the steady state solution towards which
the oscillations die (see red curve in Figure~\ref{appfig1}(a)). As in the
Brusselator model, single snapshots of the fixed point already show that the
heterogeneity of the steady state increases with the diffusion (see green
crosses, blue pluses and red circles in Figure~\ref{appfig1}(b)).  The analysis
of the dispersion of the fixed point, explained in detail in the paper,
unveils again that the steady state corresponds to a diffusion-induced
modification of the single-node solution,
$(x^{(0)},y^{(0)})\simeq(0.27,0.27)$. Indeed, the black crosses in
Figure~\ref{appfig1}(a) and black squares in \ref{appfig1}(b) indicate the
goodness of the dispersion equation to obtain the underlying fixed point.
Finally, stability analysis reveals again a Hopf bifurcation towards the
amplitude death and its posterior restoration, in a similar fashion as for
the Brusselator system (see Figure~\ref{appfig1}(c) and~\ref{appfig1}(d)).
Overall, the  observed phenomenology proves ubiquitous despite the strong
differences on the  formulation of the reactive terms for both systems.

\subsection{Topology and parameter modification}

\begin{figure*}[t]
        \includegraphics[width=1\textwidth]{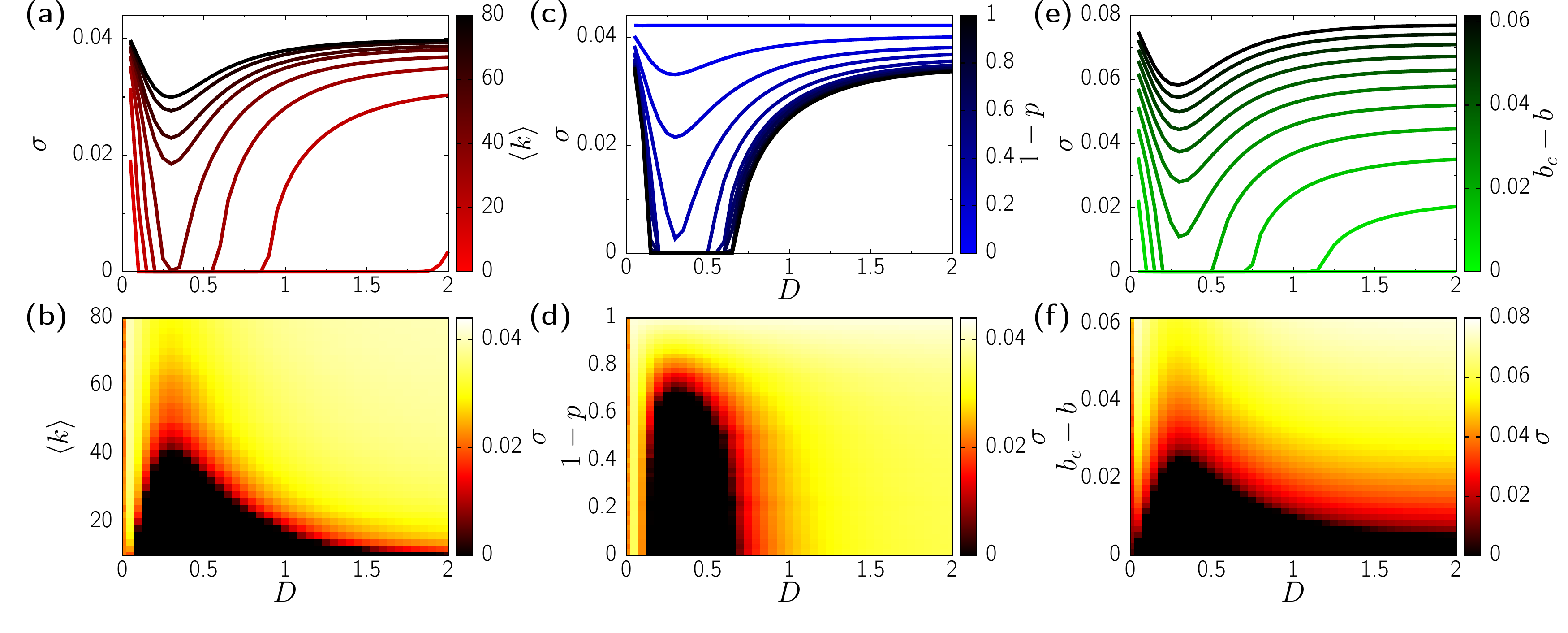} \caption{
                Amplitude death and restoration dependence on different topological and system parameters.
                Top panels show the dependence of $\sigma$ on $D$ where each line corresponds to a specific parameter value. Bottom panels show heatmaps of the amplitude upon tuning the diffusion and the parameter. System parameters are $a=1$, $d=0.1$, and $b=0.25$ , except in panels (e,f) where $b$ is modified as indicated.
                (a,b) Dependence of the oscillation amplitude $\sigma$ on the diffusion $D$
                in a range of ER networks with different average degree $\langle k\rangle$.
                (c,d) Dependence of the oscillation amplitude $\sigma$ on the diffusion $D$
                for a range of Watts-Strogatz networkss with rewiring probability $p\in[0,1]$.
                (e,f) Dependence of the oscillation amplitude $\sigma$ on the diffusion $D$
                for different values of the system parameter $b$ in a single ER network with $\langle k\rangle=20$.}
        \label{appfig2}
\end{figure*}

Figure~\ref{appfig2} shows the same instances reported in the paper where
topology and parameter modification vary the scenario depicted in the
previous section.  In particular, an increase of the density in the network
(see Figures~\ref{appfig2}(a,b)) reduces the region of steadiness, until both
critical points collapse. The same occurs for the WS networks upon
increasing the regularity of the networks (see Figures~\ref{appfig2}(c,d)).
Finally, the same results are obtained by keeping the topology fixed and
tuning the system parameter $b$. Closer to the single-node  bifurcation
point $b_c$, the coupling rapidly tends to kill the periodic behavior,
whereas increasing $b$ finally vanishes the AD phenomena (see Figures
\ref{appfig2}(e,f)).

\subsection{Heterogeneous mean-field approach}
\label{app7}

The heterogeneous mean-field approach also
provides a good agreement with the original simulations, although here the self-consistency
of the problem needs further numerical effort to be solved, since the nonlinearities in the formulation of the system limit the analytical results.\\

 Let us recover system~(\ref{eq1}) but
 we now substitute the elements of the adjacency matrix $a_{ij}$ by the annealed network approximation,
 \begin{equation*}
 a_{ij}=\frac{k_i k_j}{\langle k\rangle N}\;.
 \end{equation*}
 After applying this transformation in the coupling terms of Eq. (\ref{eq1})
 and doing some straightforward calculations one obtains the mean-field system
 \begin{equation}\label{appeq:meanfield}
 \begin{cases}
 \dot x_i=&f(x_i,y_i)+D\left( \tilde k_i \overline x-x_i \right) \\[10pt]
 \dot y_i=&g(x_i,y_i)+D\left( \tilde k_i \overline y-y_i \right)\\
 \end{cases}
 \end{equation}
 where $\tilde k_i:=k_i/\sum_{j=1}^N k_j$, and $\overline x=\frac{1}{N}\sum_{j=1}^N x_j$ and
 $\overline y=\frac{1}{N}\sum_{j=1}^N y_j$ are the mean-field activities of $x$ and $y$ variables.
Here, in order to solve for the fixed point, $\dot x_i=0$ and $\dot y_i=0$ one needs to
find the roots of a 5th order polynomial, thus, unlike the Brusselator model,
the expressions of $x_i$ and $y_i$ cannot be explicitly obtained. Therefore,
given a value of $\overline x$ and $\overline y$ we obtain the fixed points $\left(x_i^{(0)},y_i^{(0)})\right)$ for $i=1,\dots,N$ numerically.
Proceeding this way, one can approximate the values of $\overline x$ and $\overline y$
up to the desired accuracy.\\

\begin{figure*}[t]
  \includegraphics[width=1\textwidth]{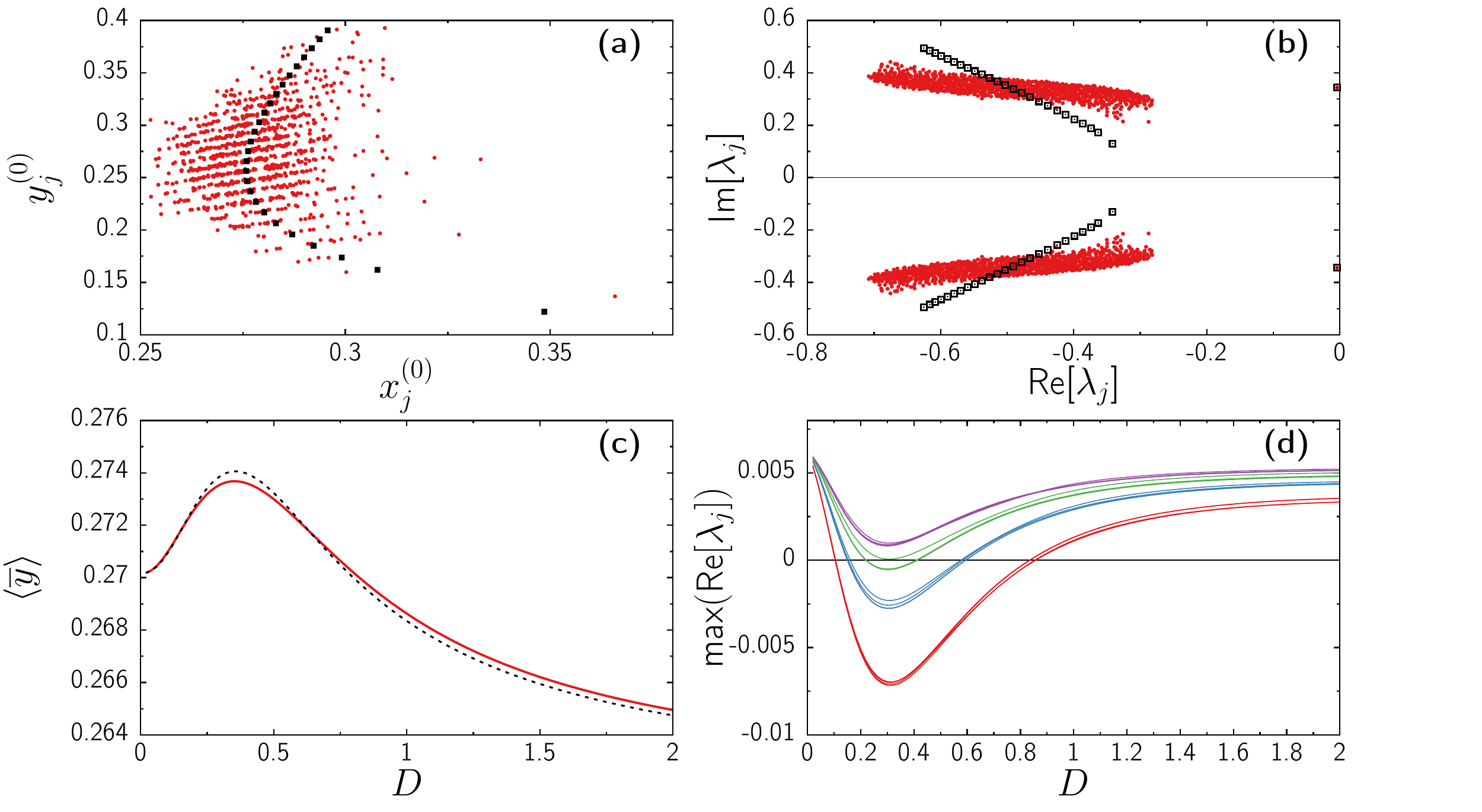} \caption{
    Mean-field analysis of the Holling-Tanner model with $a=1$, $b=0.25$,
    and $d=0.1$.  (a) Red circles correspond to the fixed point for $D=2$ as
    obtained from the numerical solution for the fixed point on the original
    system. Black squares correspond to the mean-field result for the same
    value of $D$.  (b) Red circles correspond to the eigenvalues of the
    fixed point for $D=2$, whereas open black square indicate the spectra
    resulting from the mean-field reduction.  (c) Average activity of the
    fixed point obtained from directly solving the fixed point system of
    equations.  (red continuous curve), and from the mean-field reduction
    (black dashed curve).  (d) Largest eigenvalue's real part for the
    mean-field solution with ER networks with average degree $\langle
    k\rangle =30$ (red), 30 (blue), 40 (green), and 50 (purple).  For each
    set of networks each line denotes a different network size. From top to
    bottom, $N=10^3$, $10^4$, $10^5$, and $10^6$.  }
        \label{appfig3}
\end{figure*}

\begin{figure*}[t]
  \includegraphics[width=1\textwidth]{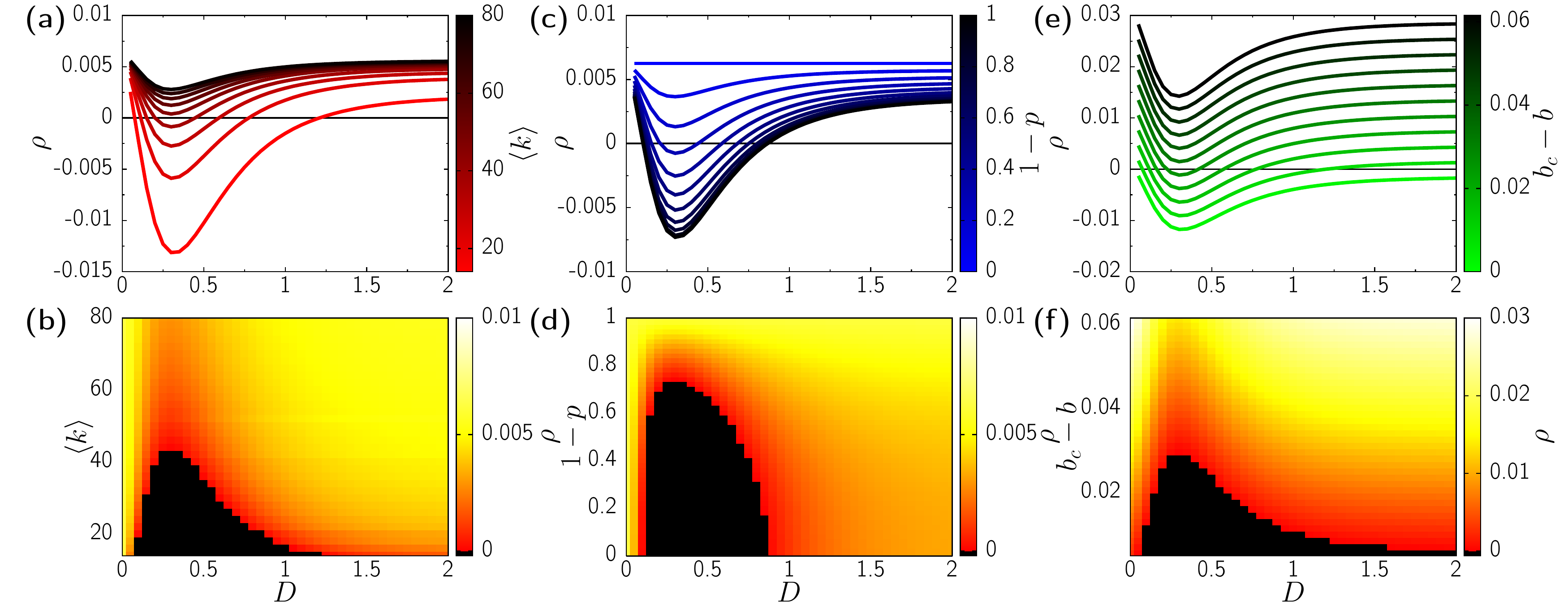}
  \caption{ Results of the HMF analysis for the Holling-Tanner model with
    $a=1$, $b=0.25$, and $d=0.1$.  Top panels show the dependence of $\rho$
    on $D$ where each line corresponds to a specific parameter value. Bottom
    panels show heatmaps of the amplitude upon tuning the diffusion and the
    parameter. System parameters are $a=1$, $d=0.1$, and $b=0.25$ , except
    in panels (e,f) where $b$ is modified as indicated.  (a,b) Dependence of
    $\rho$ on the diffusion $D$ in a range of ER networks with different
    average degree $\overline k$.  (c,d) Dependence of the oscillation
    amplitude $\rho$ on the diffusion $D$ for a range of Watts-Strogatz
  networkss with rewiring probability $p\in[0,1]$.  (e,f) Dependence of the
oscillation amplitude $\rho$ on the diffusion $D$ for different values of
the system parameter $b$ in a single ER network with $\langle k\rangle=20$.
}
        \label{appfig4}
\end{figure*}

Figure~\ref{appfig3} shows the results corresponding to this analysis.  Again,
the heterogeneous mean-field approach provides a good representation of the
underlying scenario with ER. Therefore, we implement this dimensionality
reduction approach to study the amplitude death and restoration in large
networks.  Again, as it happens in the Brusselator system, increasing the
network size does not effect the observed phenomenology.  Therefore, we
finally repeat the analysis depicted in Figure~\ref{appfig2} but using networks
with $N=10^6$ nodes analyzed through HMF, using the maximum of the largest
eigenvalue of the system,
\begin{equation*}
\rho:=\max(Re[\lambda_j])
\end{equation*}
as a proxy to identify whether there oscillations or not. Results in
Figure~\ref{appfig4}.


\section{Supplementary material for the Brusselator system}

\begin{figure*}[t]
        \includegraphics[width=1\textwidth]{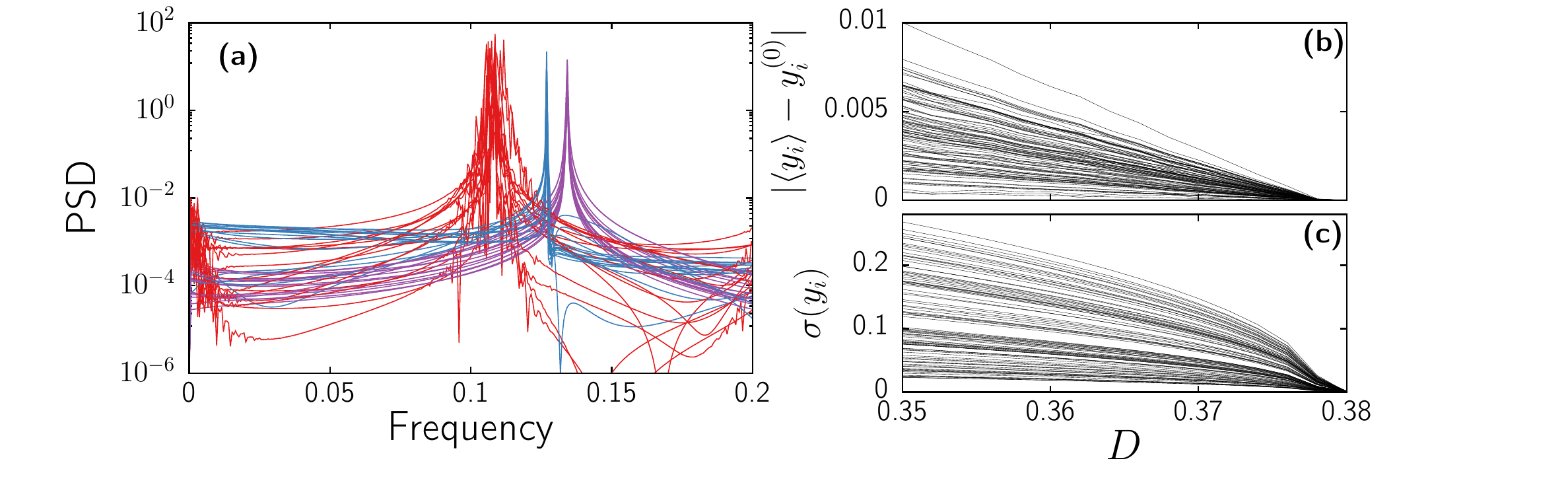} \caption{
                (a) Power Spectral Density of the time series corresponding to figure 2(a) of the main paper.
                Red curves correspond to $D=0.02$, blue corresponds to $D=0.3$, and purple to $3.5$. Results corresponding to 15 different nodes.
                For each value of the diffusion the PSD of the 15 nodes depicted in figure 2(a) of the paper are displayed.
                (b) Distance between the focus of each node limit-cycle, $\langle y_j\rangle$ and the underlying fixed point $y_j^{(0)}$
                for different values of $D$. Only results of 100 nodes shown.
                (c) Amplitude of each node evolution $\sigma(y_j)$ for different values of $D$. Only results of 100 shown.
        }
        \label{appfigSM}
\end{figure*}

\begin{figure*}[t]
        \includegraphics[width=1\textwidth]{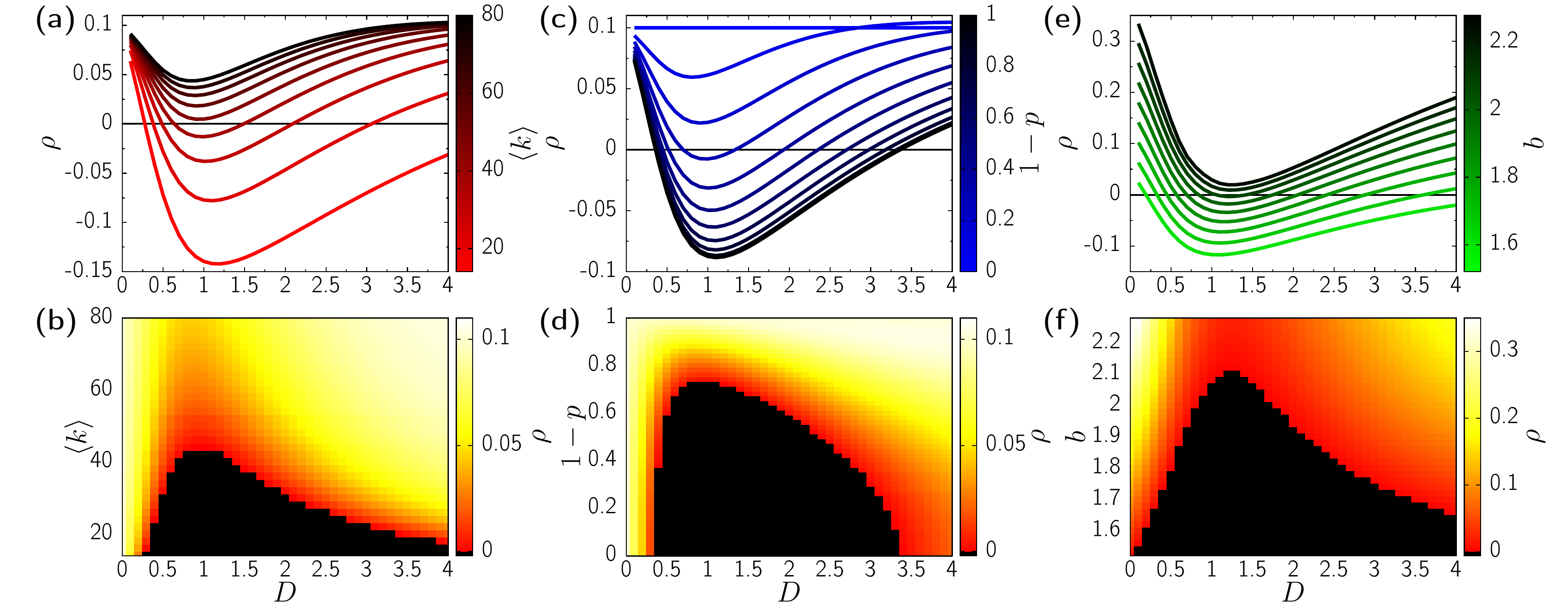} \caption{
                Results of the HMF analysis for the Brusselator model.
                Top panels show the dependence of $\rho$ on $D$ where each line corresponds to a specific parameter value. Bottom panels show heatmaps of the amplitude upon tuning the diffusion and the parameter. System parameters are $a=1$, $d=0.1$, and $b=0.25$ , except in panels (e,f) where $b$ is modified as indicated. Recall that $\rho:=\max_j(\text{Re}[\lambda_j])$.
                (a,b) Dependence of $\rho$ on the diffusion $D$
                in a range of ER networks with different average degree $\overline k$.
                (c,d) Dependence of the oscillation amplitude $\rho$ on the diffusion $D$
                for a range of Watts-Strogatz networks with rewiring probability $p\in[0,1]$. .
                (e,f) Dependence of the oscillation amplitude $\rho$ on the diffusion $D$
                for different values of the system parameter $b$ in a single ER network with $\langle k\rangle=20$. }
        \label{appfig7}
\end{figure*}

Figure~\ref{appfigSM}(a) shows the Power Spectral Density corresponding to time
series depicted in Figure~2 of the main paper for diffusion values $D=0.02$,
$D=0.3$, and $D=3.5$. For $D=0.02$ the spectra corresponding to different
nodes are irregular, but the peak is only slightly shifted around a similar
value, thus the global dynamics corresponds to irregular fluctuations
evolving at similar frequencies.  For $D=0.3$ and $D=3.5$ the spectra shows
a sharp peak, indicating that the frequency entrainment is total.  In
Figure~\ref{appfigSM}(b) we show how the centroid of each node's limit-cycle
approaches the (unstable) fixed point $y_j^{(0)}$ as $D$ increases up to
$D_1$. The centroid are computed as the time-averaged value over a long
enough time series, $\langle y_j\rangle$. Figure \ref{appfigSM}(c) shows the
decrease of the amplitude of each node's oscillations until it vanishes at
$D_1$.  Finally, figure \ref{appfig7} displays the results of the HMF analysis
for networks with $N=10^6$ analogous to figure 5 of the main paper.

\bibliography{references}

\section*{Acknowledgements}

We acknowledge financial support from the Spanish MINECO, under Projects No.
FIS2016-76830-C2-1-P and No. FIS2016-76830-C2-2-P. R.P.-S. acknowledges
additional financial support from ICREA Academia, funded by the Generalitat
de Catalunya.

\end{document}